

\input phyzzx
\input epsf
\catcode`\@=11 

\def\mychapter#1{
   \par \penalty-300 \vskip\chapterskip
   \chapterreset \phantom{-} \vskip .6in \titlestyle{CHAPTER \chapterlabel.}
    \vskip .1in  \titlestyle{#1} \vskip .2in
   \nobreak\vskip\headskip \penalty 30000
	\immediate\write16{ }
	\immediate\write16{Chapter \chapterlabel:  #1}
	\immediate\write16{ }
        \count2=\chapterlabel
   {\pr@tect\wlog{\string\chapter\space \chapterlabel}}}

\def\mysection#1{
   \par \penalty-300 \vskip\chapterskip
   \chapterreset \phantom{-} \vskip .6in \titlestyle{SECTION \chapterlabel.}
    \vskip .1in  \titlestyle{#1} \vskip .2in
   \nobreak\vskip\headskip \penalty 30000
	\immediate\write16{ }
	\immediate\write16{Chapter \chapterlabel:  #1}
	\immediate\write16{ }
        \count2=\chapterlabel
   {\pr@tect\wlog{\string\chapter\space \chapterlabel}}}

\catcode`\@=12 

\def\ev{{\rm e\kern-.1100em V}}
\def\gev{{\rm G\kern-.1000em \ev}}
\def\tev{{\rm T\kern-.1000em \ev}}
\def\U#1{U\kern-.25em \left( #1 \right) }
\def\su#1{S\U #1 }
\def\up#1{\kern-0.45em\raise0.4em\hbox{$\scriptstyle #1 $}\hskip0.29em}
\def\sqr#1#2{{\vcenter{\vbox{\hrule height.#2pt
        \hbox{\vrule width.#2pt height#1pt \kern#1pt
                 \vrule width.#2pt}
             \hrule height.#2pt}}}}

 \font\bigboldiii=cmbx10 scaled\magstep3

 \def\dcal{{\cal D}}
 \def\lcal{{\cal L}}
 \def\ocal{{\cal O}}
 \def\ui{U(1)}
 \def\and{{\it\&}}
 \def\half{{1\over2}}
 \def\lowti#1{_{{\rm #1 }}}
 \def\tr{ \hbox{tr}}
 \def\quarter{{1\over4}}
 \def\gesim{\,{\raise-3pt\hbox{$\sim$}}\!\!\!\!\!{\raise2pt\hbox{$>$}}\,}
 \def\lesim{\,{\raise-3pt\hbox{$\sim$}}\!\!\!\!\!{\raise2pt\hbox{$<$}}\,}
 \def\BB{{\bf B}}
 \def\DD{{\bf D}}
 \def\WW{{\bf W}}

\def\pl{{\phi_l}}
\def\ph{{\phi_h}}
\def\fl{{\psi_l}}
\def\fh{{\psi_h}}
\def\Dslash{\not\!\! D}

\FIG\fourfermi{The contribution of two types of underlying physics to the
 four-fermi operator. (a) A heavy gauge boson leads to a coupling constant of
 order one.  (b) A heavy fermion comes in only at the one-loop level and leads
 to a coupling constant of order $1/16\pi^2$}
\FIG\allvertices{These are the possible vertices in an unconstrained field
 theory.  Each line may be heavy or light.  Each dashed line may be a scalar or
 a vector.}
\FIG\allgraphs{These are tree graphs which are suppressed by at most
 $1/\Lambda^2$.  Internal lines are all heavy, and external lines are all
 light.  Each dashed line may be a scalar or a vector.}
\FIG\dimeight{A dimension-eight operator induced at tree level.}

 \def\NP{{\sl Nucl. Phys.} }
 \def\PL{{\sl Phys. Lett.} }
 \def\PR{{\sl Phys. Rev.} }
 \def\PRL{{\sl Phys. Rev. Lett.} }

\REF\einhorn{
 M.B. Einhorn and J. Wudka, preprint UM-TH-92-25;
 NSF-ITP-92-01, {\sl Proceedings of the Workshop on
 Electroweak Symmetry Breaking,} Hiroshima, Nov. 12-15 19091, Singapore:
 World Scientific, to be published.}
\REF\der{
 A. De R\'ujula, M.B. Gavela, P. Hernandez, and E. Maas\'o,
       \NP {\bf B384} (1992) 3.}
\REF\efflag{
 H. Georgi and H. Politzer, {\sl Phys. Rev.} {\bf D14} (1976) 1829.
 E. Witten, {\sl Nucl. Phys.} {\bf B104} (1976) 445.
 E. Witten, {\sl Nucl. Phys.} {\bf B122} (1977) 109.
 J. Collins, F. Wilczek, and A. Zee {\sl Phys. Rev.} {\bf D18} (1978) 242.
 S. Weinberg, {\sl Phys. Rev. Lett.} {\bf 43} (1979) 1566.
 F. Wilczek and A. Zee, {\sl Phys. Rev. Lett.} {\bf 43} (1979) 1571.
 Y. Kazama and Y.P. Yao, {\sl Phys. Rev.} {\bf D21} (1980) 1116.
 B. Ovrut and H. Schnitzer, {\sl Phys. Rev.} {\bf D21} (1980) 3369.
 S. Weinberg, {\sl Phys. Lett.} {\bf B91} (1980) 51.}
\REF\decou{
 T. Appelquist and J. Carazzone, {\sl Phys. Rev.} {\bf D11} (1975) 2856.
 K. Symanzik, {\sl Comm. Math. Phys.} {\bf34} (1973) 7.}
\REF\ndc{
 S. Weinberg, {\sl Physica} {\bf A96} (1979) 327.
 T. Appelquist and C. Bernard, \PR {\bf D22} (1980) 200.
 T. Appelquist and C. Bernard, \PR {\bf D23} (1981) 425.
 M. Chanowitz and M.K. Gaillard, \NP {\bf B261} (1985) 379.
 M. Chanowitz \etal, \PR {\bf D36} (1987) 1490.
 J. Gasser \etal, \NP  {\bf B307} (1988) 779.
 B. Grinstein and M. Wise, preprint HUTP-91/A015.
 T. Appelquist and J. Terning, preprint YCTP-P41-92.
 J. Bagger, S. Dawson, and G. Valencia, \PRL {\bf 67} (1991) 2256.
 M. Golden and L. Randall, \NP {\bf B361} (1991) 3.
 B. Holdom and J. Terning, \PL {\bf 247B} (1990) 88.
 B. Holdom, \PL {\bf 259B} (1991) 329, \PL {\bf 258B} (1991) 156.}
\REF\gbook{
 H. Georgi, {\sl Weak Interactions
    and Modern Particle Theory}, Benjamin / Cummings Publishing Co.,
    Menlo Park, CA, USA (1984).}
\REF\gla{
 J. Gasser and H. Leutwyler, {\sl Ann. Phys.} {\bf 158} (1984) 142.}
\REF\glb{
 J. Gasser and H. Leutwyler, \NP {\bf B250} (1985) 465,
    \NP {\bf B250} (1985) 517,
    \NP {\bf B250} (1985) 539.}
\REF\bw{W. Buchm\"uller and D. Wyler, \NP {\bf 268B} (1986) 621.}
\REF\canlag{C. Arzt, UM-TH-92-28 (hep-ph/9304230).}
\REF\hagis{
 K. Hagiwara, S. Ishihara, R. Szalapski, and D. Zeppenfeld, \hfil\break
    preprint MAD/PH/737 (March 1993).}
\REF\alt{
 G. Altarelli, R. Barbieri, and S. Jadach, \NP {\bf B369} (1992) 3.
 P. Langacker and M. Luo, {\sl Phys. Rev.} {\bf D44} (1991) 817.}
\REF\her{
 M.J. Herrero and E.R. Morales, preprint FTUAM 93/24, hep-ph/9308276.}
\REF\nda{
 A. Manohar and H. Georgi, \NP {\bf B234} (1984) 189.
 H. Georgi, \PL {298B} (1993) 187.}
\REF\lee{
 D.A. Dicus and V.S. Mathur, \PR {\bf D7} (1973) 3111.
 B.W. Lee, C. Quigg, and H.B. Thacker, \PR {\bf D16} (1977) 1519.
 M. Chanowitz and M. Gaillard, \NP {\bf B261} (1985) 379.
 W.J. Marciano and S.S.D. Willenbrock, \PR {\bf D37} (1988) 2509.
 S. Dawson and S.S.D. Willenbrock, \PRL {\bf 62} (1989) 1232.
 M. Veltman and F. Yndurain, \NP {\bf 325B} (1989) 1.
 S. Dawson and S. Willenbrock, \PR {D40} (1989) 2880.}
\REF\einb{
 M.B. Einhorn, \NP {\bf 246B} (1984) 75.}
\REF\bagger{
 J. Bagger, S. Dawson, and G. Valencia, preprint FERMILAB-PUB-92/75-T,
      hep-ph/92024211 (August 1992).}
\REF\dona{
 J.F. Donoghue, C. Ramirez, and G. Valencia, \PR {\bf D39} (1989) 1947.}
\REF\holter{
 B. Holdom and J. Terning, \PL {\bf B247} (1990) 88.}
\REF\apwu{
 T. Appelquist and G.-H. Wu, preprint YCTP-P7-93.}
\REF\qlc{
 I.J.R. Aitchison and C.M. Fraser, \PL {\bf B146} (1984) 63.
 A.A. Andrianov, \PL {\bf B157} (1985) 425.}
\REF\ecker{
 G. Ecker, J. Gasser, A. Pich, and E. de Rafael, \PL {B321} (1989) 321.}
\REF\pesta{
 M.E. Peskin and T. Takeuchi, \PRL {\bf 65} (1990) 964;
 \PR {\bf D46} (1992) 381.}
\REF\current{
 J. Alitti et. al., The UA2 Collaboration, \PL {\bf 277B} (1992) 194.
 C.P Burgess, S. Godfrey, H. K\"onig, D. London, and I. Maksymyk,
     preprint McGill-93/14, hep-ph/9307223 (June 1993).
 C. Grosse-Knetter, I. Kuss, and D. Schildknecht, preprint BI-TP 93/15,
    hep-ph/9304281 (April 1993).
 D. Choudhury, P. Roy, and R. Sinha, preprint TIFR-TH/93-08 (April 1993).
 P. Hernandez and F.J. Vegas, preprint CERN-TH 6670 (December 1992).}
\REF\hagp{
 K. Hagiwara, R.D. Peccei, and D. Zeppenfeld, \NP {\bf B282} (1987) 253.}
\REF\holdom{
 B. Holdom, \PL {\bf B258} (1991) 156.}
\REF\barb{
 G. Barbiellini et al. in {\sl Physics at LEP}, J. Ellis and R. Peccei, ed.,
    (CERN 86-02) vol.2 (1986).}
\REF\bilenky{
 M. Bilenky, J.L. Kneur, F.M. Renard, D. Schildknecht, preprint BI-TP 92/44
    (February 1993).}
\REF\kane{
 G.L. Kane, J. Vidal, and C.-P. Yuan, \PR {\bf D 39} (1989) 2617.}
\REF\barklow{
 T.L. Barklow, preprint SLAC-PUB-5808 (April 1992).}
\REF\burke{
 D.L. Burke in {\sl Gauge Bosons and Heavy Quarks}, J. Hawthorne, ed.,
    Proceedings of the 18th SLAC Summer Institute on Particle Physics,
    preprint SLAC-REPORT-378, 1991.}
\REF\falk{
 A.F. Falk, M. Luke, and E.H. Simmons, \NP {\bf B365} (1991) 523.}
\REF\long{
 A. Longhitano, {\sl Nucl. Phys.} {\bf B188} (1981) 118.}
\REF\bess{See e.g., R. Casalbuoni et.al., in
``2nd International Workshop on Physics and Experiments
with Linear e+ e- Colliders, Waikoloa, HI, 26-30 Apr 1993,"
F. Harris et al. (eds.), vol. II.  Singapore: World Scientific,
1993,(hep-ph/9306262) and references therein.}




\setbox21=\vbox {
\hsize=2truein
\epsfxsize=\hsize
\epsffile{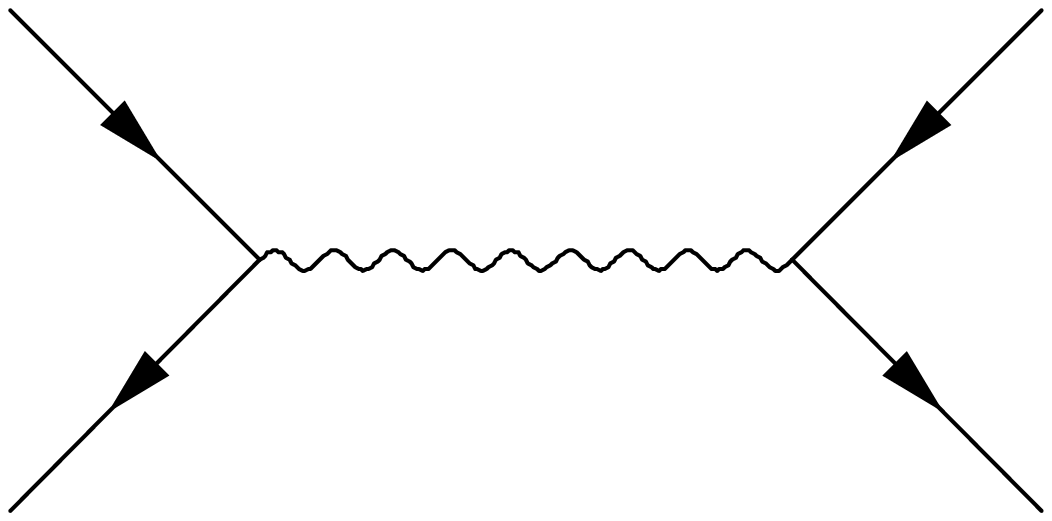}}

\setbox22=\vbox {
\epsfysize=\ht21
\epsffile{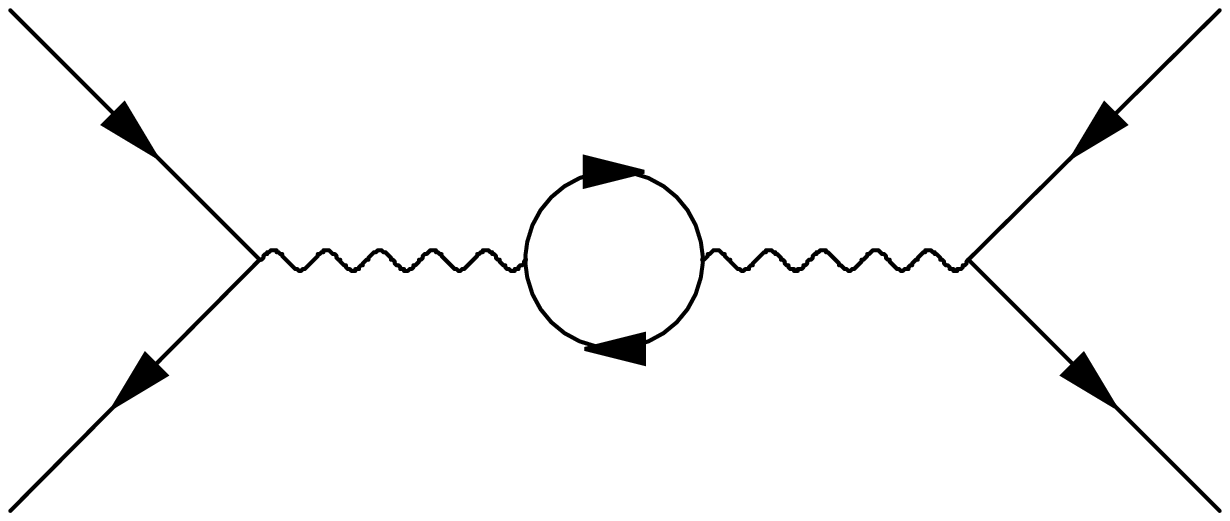}}

\singlespace
\setbox23=
\vbox {     \line { \hfill \box21 \hfill \box22 \hfill }
            \line {\ninerm \hfill(a) \hfill (b) \hfill}
{\twelverm Figure~\fourfermi}
{\ninerm The contributions to the
 four-fermion operator of two types of underlying physics. (a) A heavy gauge
boson leads to a effective coupling of
 order one.  (b) A heavy fermion comes in only at the one-loop level and leads
 to an effective coupling of order $1/16\pi^2$.}}
\doublespace


\setbox31=\vbox {
\hsize=1truein
\epsfxsize=\hsize
\epsffile{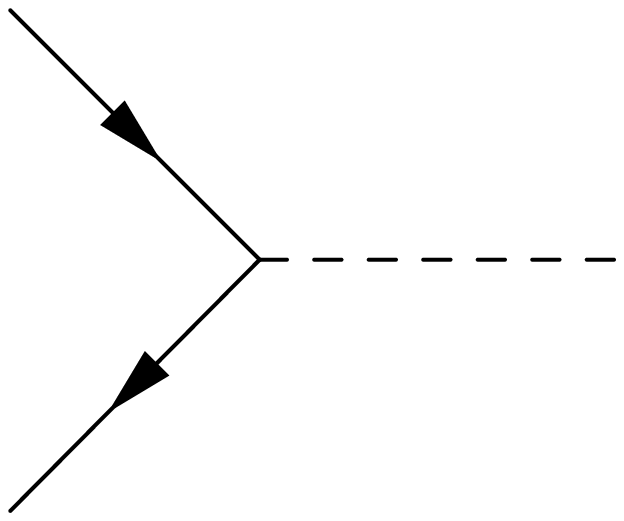}}

\setbox32=\vbox {
\epsfysize=\ht31
\epsffile{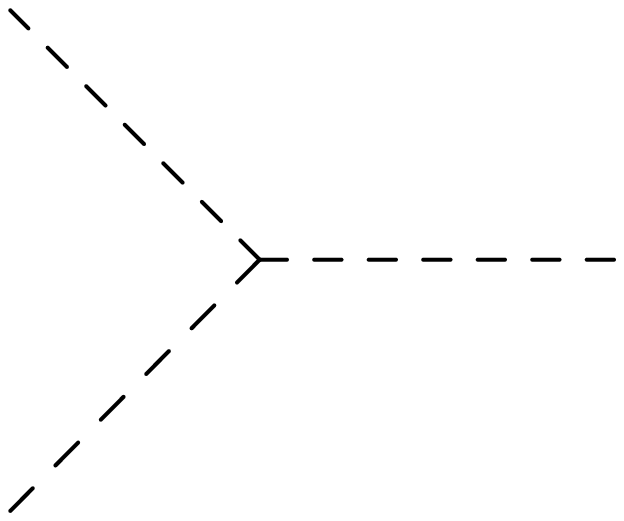}}

\setbox33=\vbox {
\epsfysize=\ht31
\epsffile{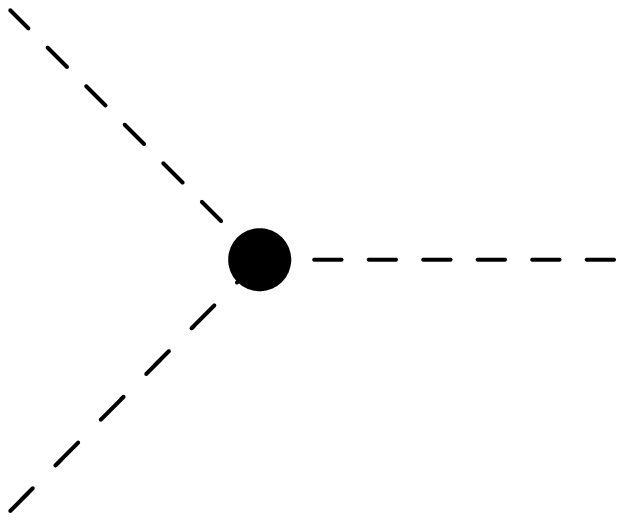}}

\setbox34=\vbox {
\epsfysize=\ht31
\epsffile{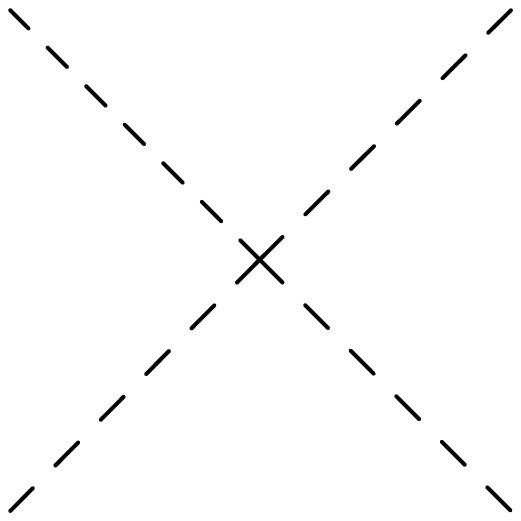}}

\singlespace
\setbox35 = \vbox {
\line{ \hfill \box31 \hfill \box32 \hfill \box33 \hfill \box34 \hfill}
\line{ \hskip2.1cm (a) \hskip3.0cm (b) \hskip3.0cm (c) \hskip3.2cm (d) \hfill}
{\twelverm Figure~\allvertices}
{\ninerm These are all possible relevant vertices in an unconstrained field
theory. Each dashed line may be a scalar or a vector.  The dot represents a
coupling constant proportional to $\Lambda$, the heavy scale.}}
\doublespace


\setbox1=\vbox {
\hsize=1.5truein
\epsfxsize=\hsize
\epsffile{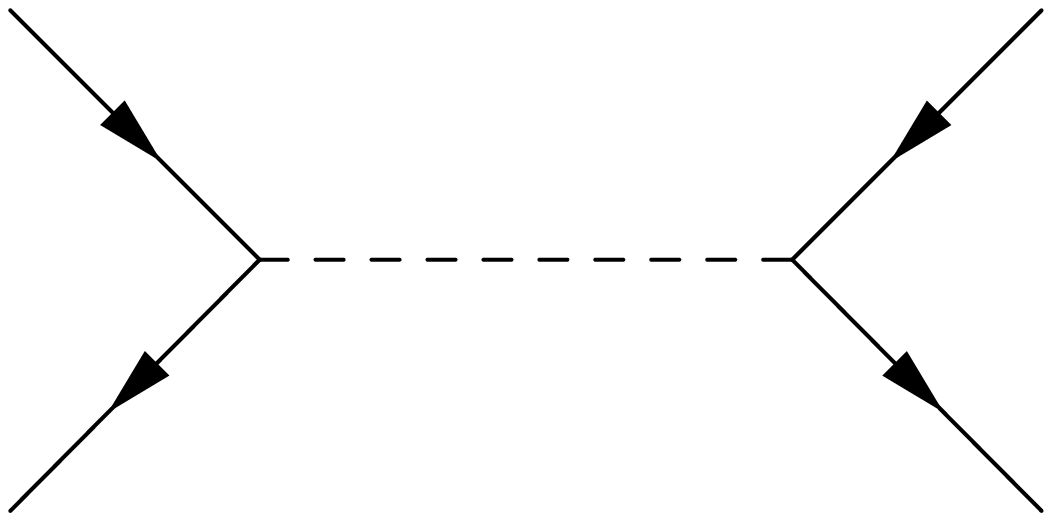}}

\setbox2=\vbox {
\epsfysize=\ht1
\epsffile{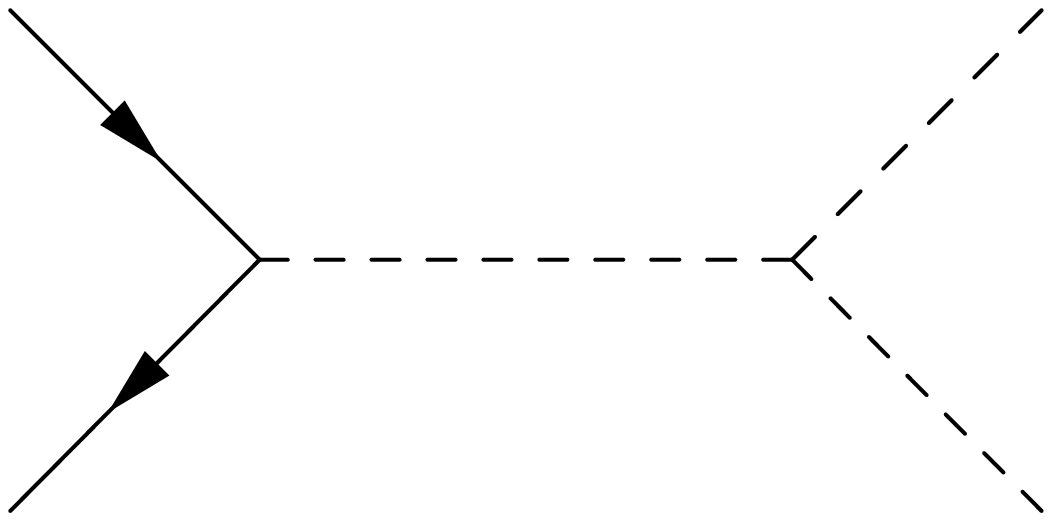}}

\setbox3=\vbox {
\epsfysize=\ht1
\epsffile{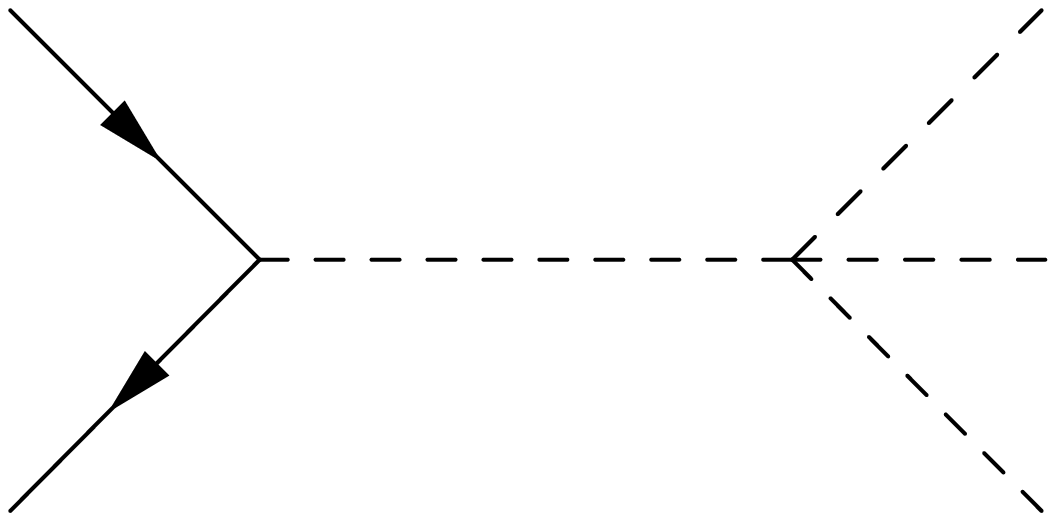}}

\setbox4=\vbox {
\epsfysize=\ht1
\epsffile{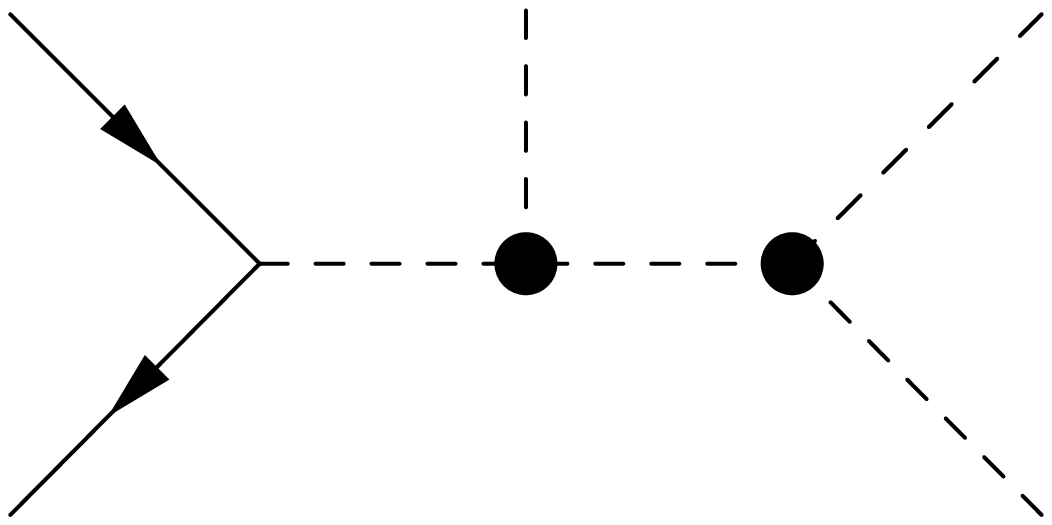}}

\setbox5=\vbox {
\epsfysize=\ht1
\epsffile{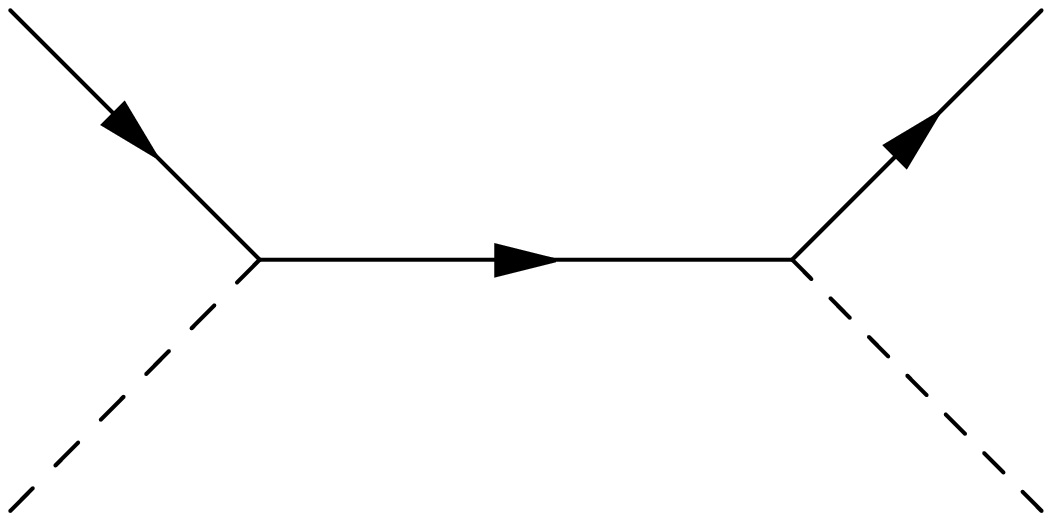}}

\setbox6=\vbox {
\epsfysize=\ht1
\epsffile{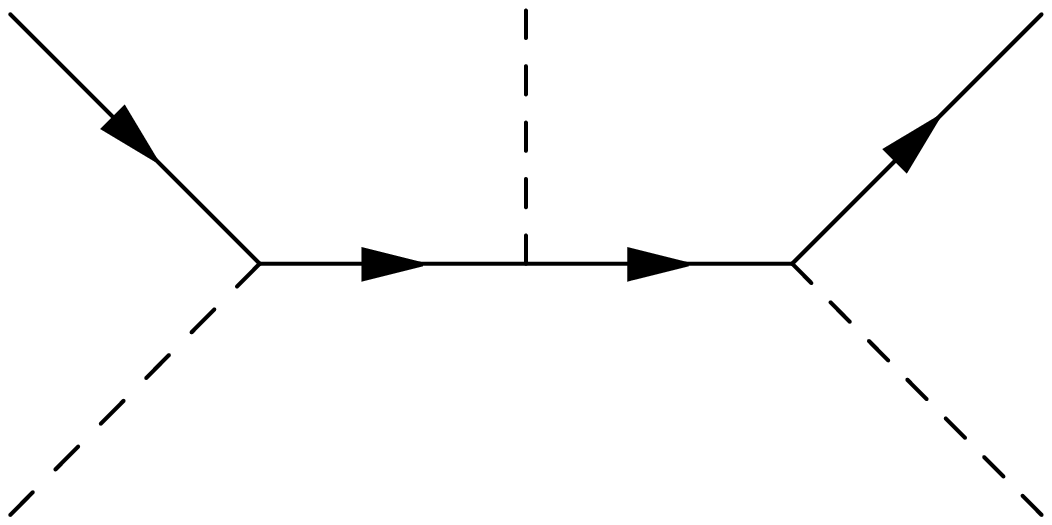}}

\setbox7=\vbox {
\epsfysize=\ht1
\epsffile{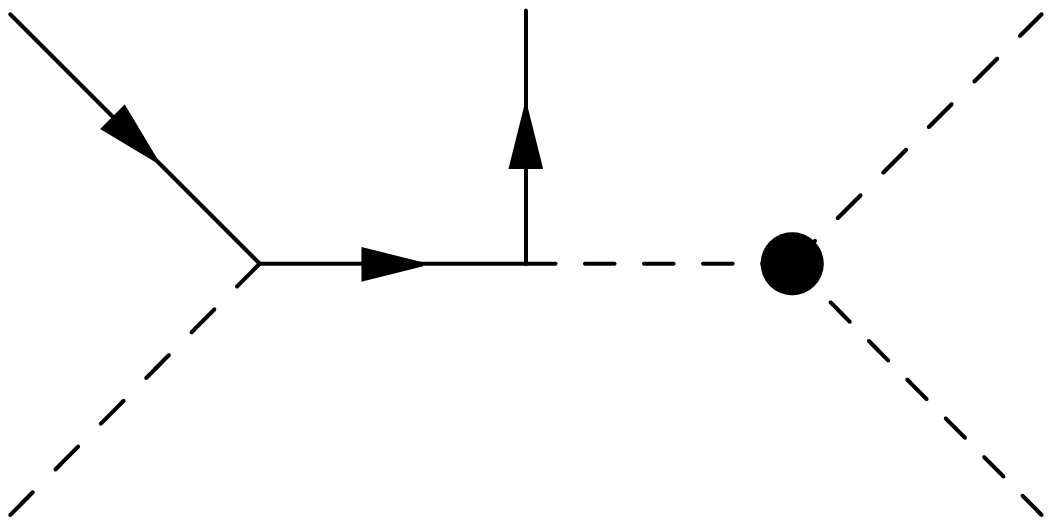}}

\setbox8=\vbox {
\epsfysize=\ht1
\epsffile{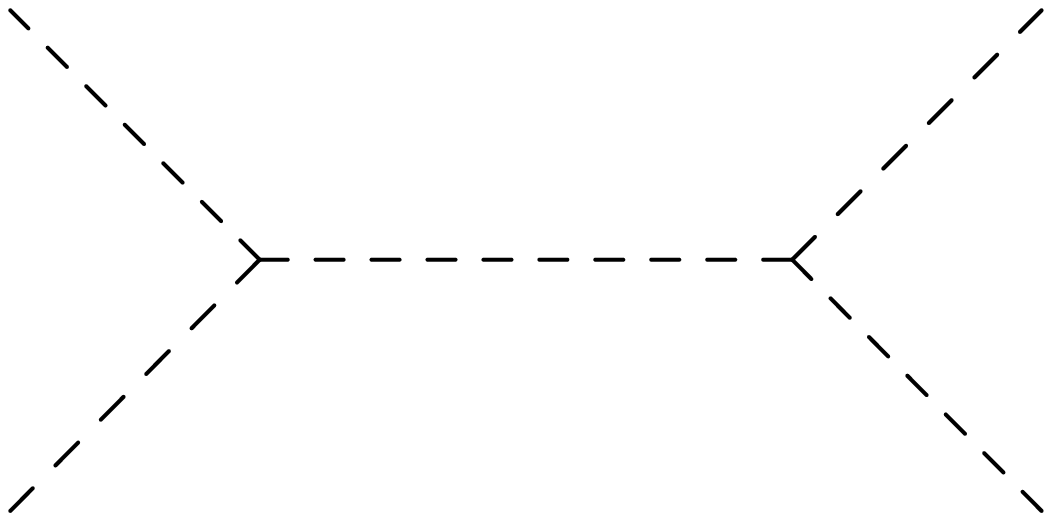}}

\setbox9=\vbox {
\epsfysize=\ht1
\epsffile{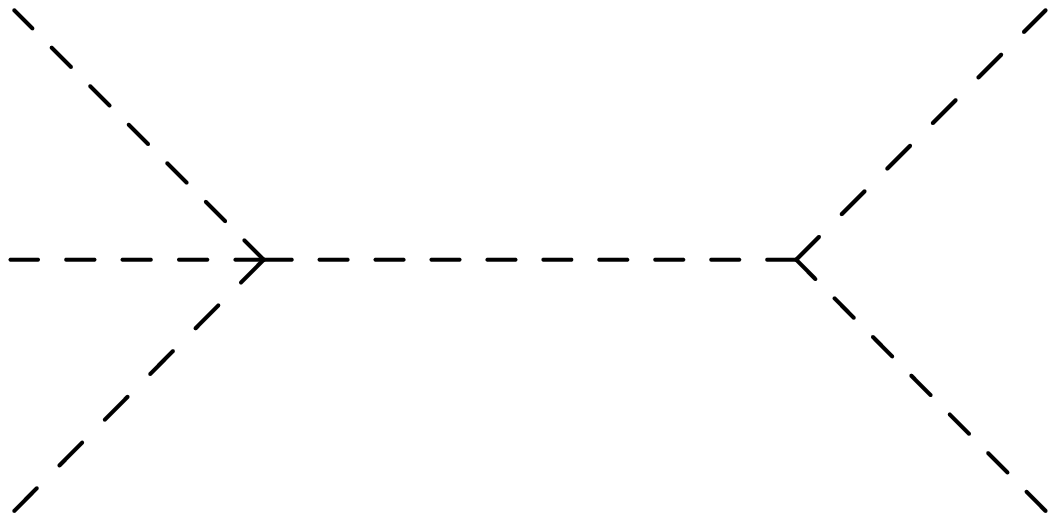}}

\setbox10=\vbox {
\epsfysize=\ht1
\epsffile{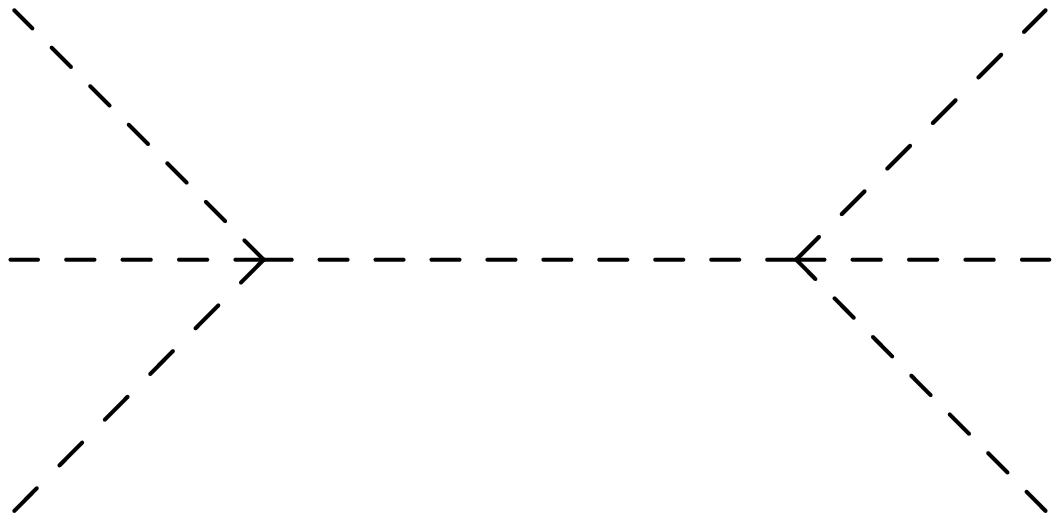}}

\setbox11=\vbox {
\epsfysize=\ht1
\epsffile{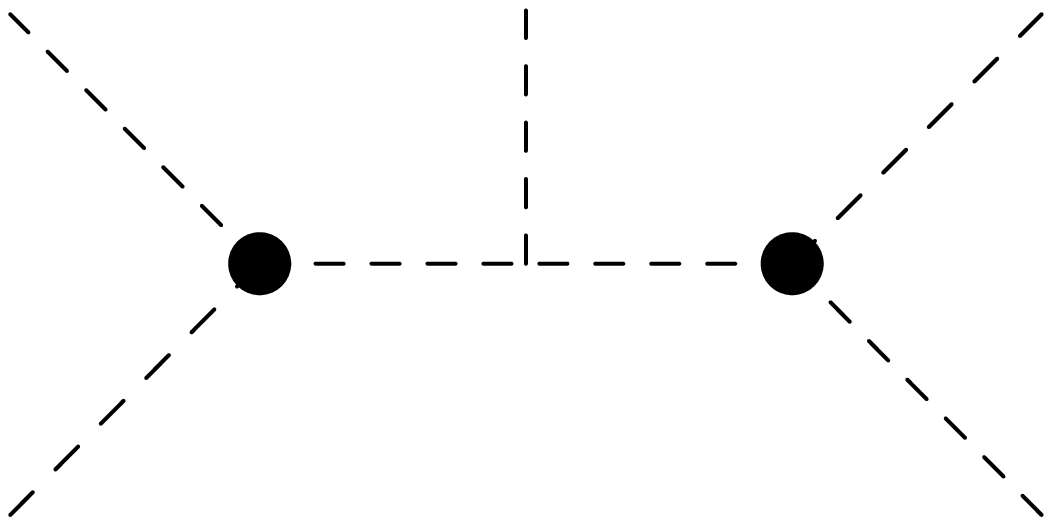}}

\setbox12=\vbox {
\epsfysize=\ht1
\epsffile{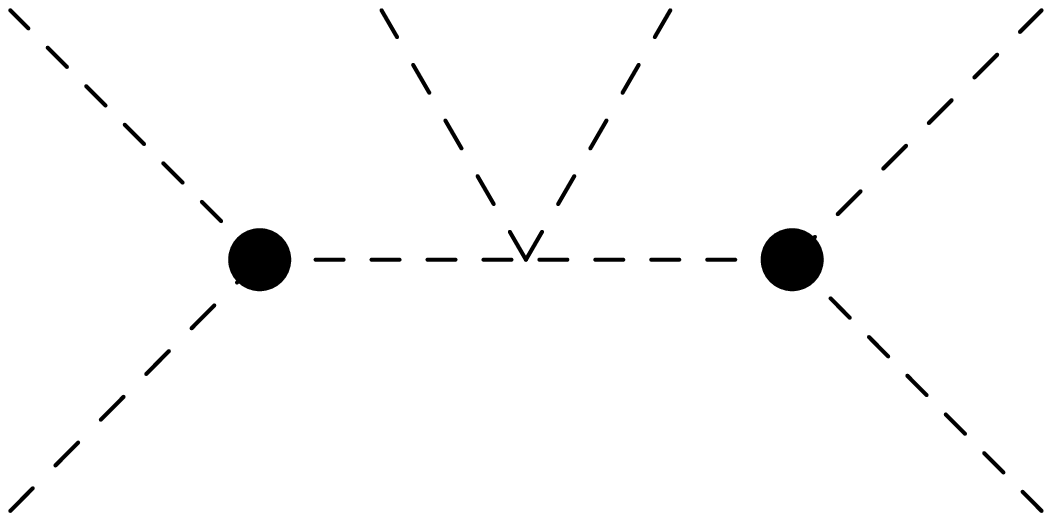}}

\setbox13=\vbox {
\epsfysize=\ht1
\epsffile{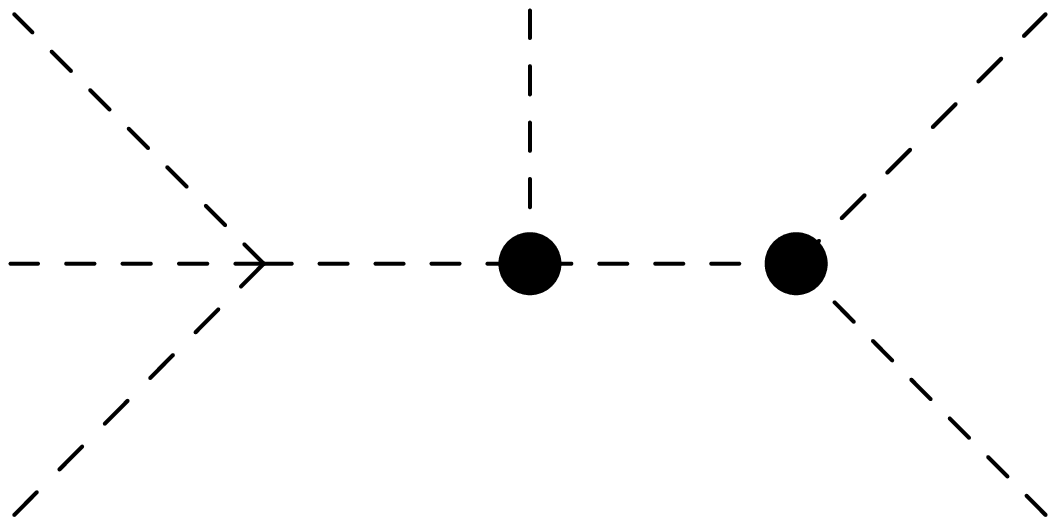}}

\setbox14=\vbox {
\epsfysize=\ht1
\epsffile{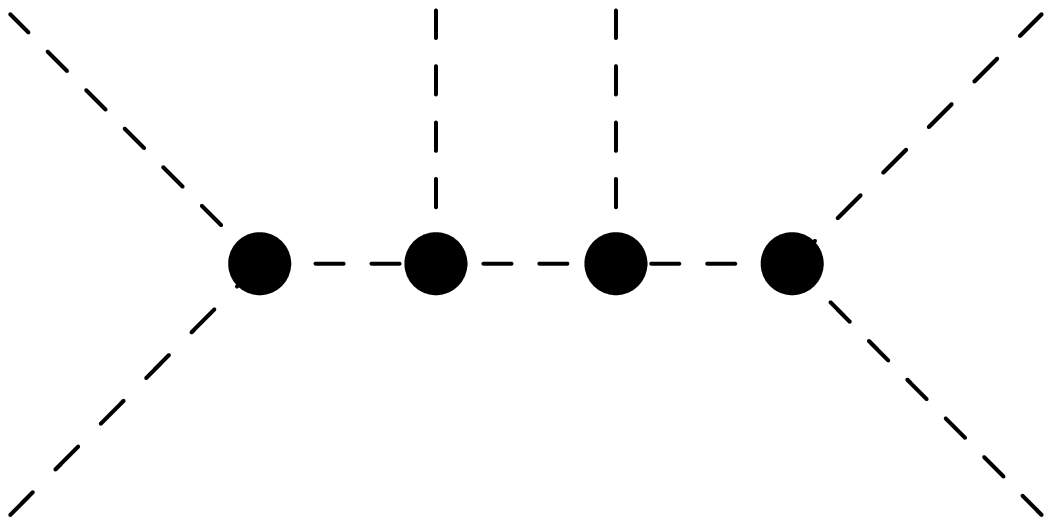}}

\setbox15=\vbox {
\epsfysize=\ht1
\epsffile{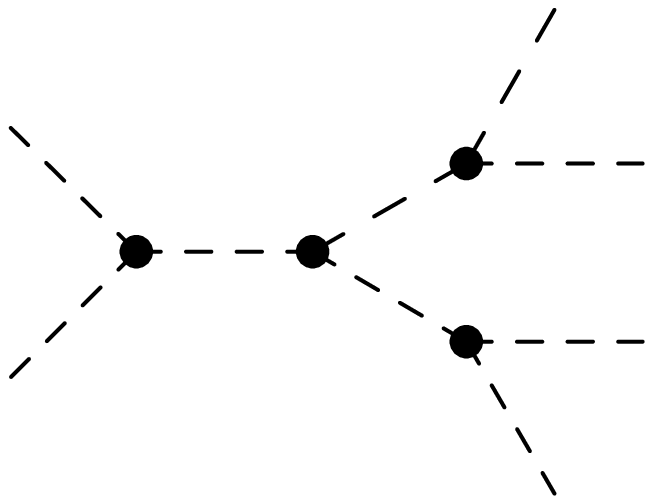}}

\singlespace
\setbox16 =
\vbox{                   \line { \hfill \box1 \hfill \box2 \hfill \box3 \hfil}
               \vskip-15pt
     \line {\ninerm \hskip1.2in (a) \hskip1.79in (b) \hskip1.77in (c) \hfill}
               \vskip15pt
                   \line { \hfill \box4 \hfill \box5 \hfill \box6 \hfil}
               \vskip-15pt
     \line {\ninerm \hskip1.2in (d) \hskip1.77in (e) \hskip1.77in (f) \hfill}
               \vskip15pt
                   \line { \hfill \box7 \hfill \box8 \hfill \box9 \hfil}
               \vskip-15pt
     \line {\ninerm \hskip1.2in (g) \hskip1.79in (h) \hskip1.77in (i) \hfill}
               \vskip15pt
                   \line { \hfill \box10 \hfill \box11 \hfill \box12 \hfil}
               \vskip-15pt
     \line {\ninerm \hskip1.2in (j) \hskip1.77in (k) \hskip1.77in (l) \hfill}
               \vskip15pt
                   \line { \hfill \box13 \hfill \box14 \hfill \box15 \hfil}
               \vskip-15pt
     \line {\ninerm \hskip1.2in (m) \hskip1.77in (n) \hskip1.77in (o) \hfill}
               \vskip15pt
{\twelverm Figure~\allgraphs}
{\ninerm
These are all possible tree graphs which are suppressed by at most
$1/\Lambda^2$, in a general field theory, with only heavy internal lines and
light external lines.  The dotted three-boson vertices must have a coupling
constant proportional to $\Lambda$, the heavy mass scale.  Each dashed line may
be a scalar or a vector.  }} \doublespace


\setbox41=\vbox {
\hsize=2truein
\epsfxsize=\hsize
\epsffile{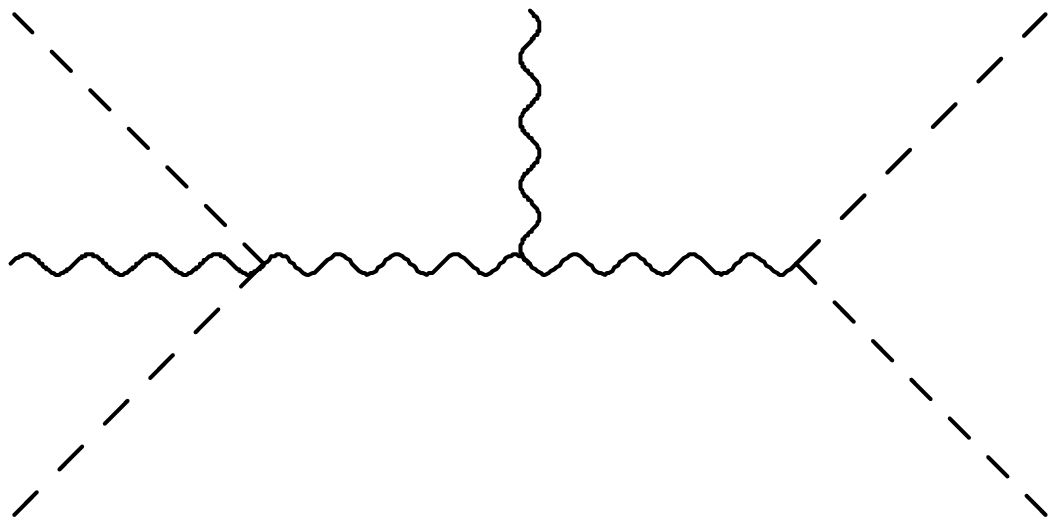}}

\setbox42=\vbox {
\epsfysize=\ht41
\epsffile{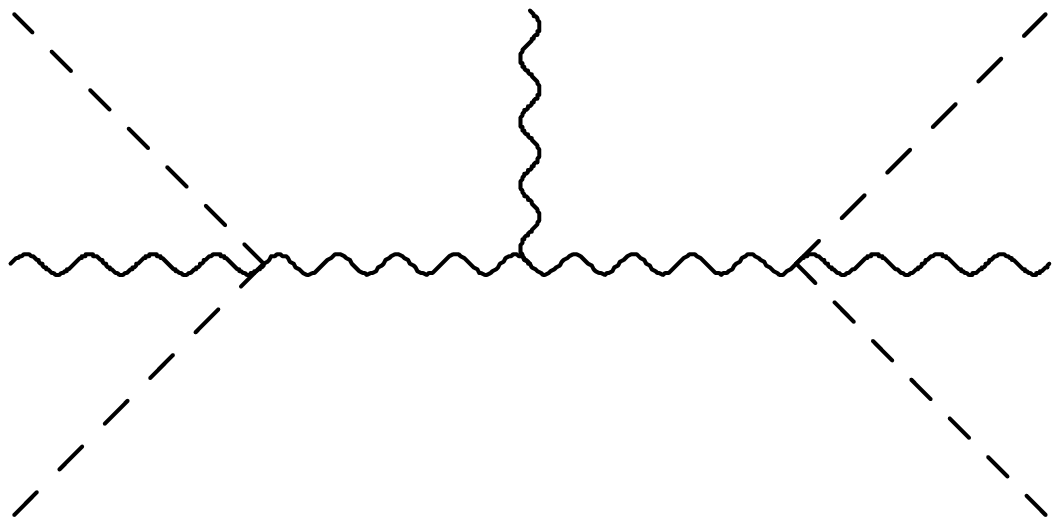}}

\singlespace
\setbox43=
\vbox {     \line { \hfill \box41 \hfill \box42 \hfill }
            \line {\ninerm \hfill(a) \hfill (b) \hfill}
{\twelverm Figure~\dimeight}
{\ninerm Tree graphs which may lead to dimension-eight effective operators.}}
\doublespace


\def\theabstract{ Effective Lagrangians can be used to parametrize the effects
of physics beyond the standard model.  Assuming the complete theory is a gauge
theory, we determine which effective operators may be generated at tree level,
and which are only generated at loop level. The latter are be suppressed by
factors of $1/ 16\pi^2$ and will therefore be quite difficult to detect.  In
particular, all operators changing the Standard-Model structure of the
triple-gauge-vector couplings fall into this category.  We also point out that
in certain cases, dimension-eight operators may be more important than
dimension-six operators.  We discuss both the linear and non-linear
representation of the Higgs sector.}

\nopubblock{
\singlespace
\rightline{$\caps UM-TH-94-15$}
\rightline{$\caps UCRHEP-125$}
\rightline{$\caps CALT-68-1932$}
\rightline{hep-ph/9405214}}
{\titlepage
\title{ {\bigboldiii
Patterns of Deviation from the Standard Model}}
\bigskip
\titlestyle{{\twelvecp C. Arzt}
{\address{{\it 452-48 Department of Physics\break
California Institute of Technology\break
Pasadena, CA 91125}} }
\medskip
\titlestyle{\twelvecp M.B. Einhorn }}
{\address{{\it Randall Laboratory of Physics\break
University of Michigan\break
Ann Arbor, MI 48109-1120 } }}
\baselineskip=12pt
\medskip
\centerline{and}
\medskip
\titlestyle{{\twelvecp J. Wudka }}
{\address{{\it University of California at Riverside\break
 Department of Physics\break
 Riverside{\rm,} California 92521{\rm--}0413{\rm,}
 U{\rm.}S{\rm.}A{\rm.} }}
\singlespace
\abstract
\theabstract
\endpage}
\doublespace

\chapter{Introduction}

The Standard Model of particle theory has been striking successful in
its experimental predictions.  Though some areas of the model have not
yet been directly tested, there are strong theoretical and
experimental arguments that the standard $\su3_C \times \su2_L \times
\ui_Y$ gauge theory provides a faithful characterization of reality at
current energies [\einhorn,\der].  It is, nonetheless, not wholly
satisfactory. The mass-generating sector, with its fundamental scalar
Higgs particle, has serious theoretical deficiencies.  Furthermore,
the large number of ``elementary" constants and particles is
aesthetically unappealing.  Attempts to solve one or another of these
problems have led to the the creation of ``high-energy" models
putatively describing physics beyond the Standard Model - GUTs,
supersymmetry, technicolor, and a host of others.

One goal of the next generation of colliders is to provide real
evidence for one of these models, or for one as yet undreamt of.
Detection of a sneutrino or a technifermion would, of course, give the
game away.  But even if we are not so lucky, we can hope to observe
the virtual effects of new physics on the interactions of
Standard-Model particles.  After all, the effects of the electroweak
theory were first seen as virtual effects beyond the
then-standard-model QED.  Believers in one or another of the
high-energy models can calculate the effects of their favorite model
on any given process.  But the profusion of possibilities, and the
chance that none now known is correct, push one towards a
model-independent analysis.

The framework for such an analysis is readily available in the form of
an effective Lagrangian approach [\efflag,\decou].  Assume for a
moment we knew the correct high-energy theory.  Integrate out the
heavy (compared to Standard-Model) fields and the high-energy modes of
the light (i.e. Standard-Model) fields.  What remains is an infinite
tower of local terms of higher and higher dimension, suppressed by
higher and higher powers of some heavy mass.  Each term is an
$SU(3)_C
\times SU(2)_L \times U(1)_Y$ symmetric effective operator built from
Standard-Model fields.  This is true {\it regardless} of the
particular high-energy model from which we started (assuming only that
it respects the gauge symmetries of the Standard Model).  Two
different theories will generally produce the same operators; the
difference between the two will be reflected in the different
numerical factors and coupling constants associated with each
operator.  The effective Lagrangian we will use is a set of such
operators, each with an arbitrary effective coupling; it can serve to
represent the effects of {\it any} high-energy theory.  What remains
to be determined are the coupling constants that distinguish one
possible high-energy scenario from another.

The goal of new experiments, then, in the absence of new-particle production,
is to determine the values of the effective coupling constants. Any experiment
which deviates from the Standard-Model prediction will require a non-zero
value for some effective coupling constant.  All this assumes, of course,
that the Standard Model is the complete description of low energy physics.
If there are any other light fields (with masses similar to those of the
Standard Model), these would have to be included in the set of fields from
which effective operators are constructed.

There have been many attempts to determine the sensitivity of upcoming
experiments to various effective operators; these may be stated in terms of
limits on the associated coupling constants.  If we are to look for
deviations from Standard-Model predictions, it is important to get a
feeling for the orders of magnitude we expect for the effective coupling
constants.  Some may be far larger than others, indicating those processes
most sensitive to new physics.  Since the strength of an effective operator
is determined by the strength of the corresponding Green's function in the
high-energy theory, operators corresponding to tree graphs in the full
theory will make their presence known much more strongly than will
operators corresponding to loop graphs, at least
whenever the heavy physics is weakly interacting.
Experiments which can only produce
limits far larger than the expected size of these couplings are
phenomenologically uninteresting.  In this paper we will show that there
are grounds to expect deviations from the Standard Model to show up more
strongly in some operators (and therefore in some processes) than in
others.

In section two we will consider a general high-energy gauge theory which, at
low energies, decouples from the Standard Model.  We divide all of the
dimension-six effective operators into those which may be generated at tree
level, and those which can only be produced by loops in the full theory.
In section three, we show that for some vertices (and specifically for the
trilinear gauge vertices) the effects of dimension-eight operators may be
larger than those of dimension-six operators.  In section four we examine the
range of effective couplings expected in a range of more general theories.  In
section five we look more closely at the trilinear gauge vertices,
comparing their expected coupling strengths to those that are observable at
present and future colliders.  We summarize our results in section six.

\chapter{Decoupling Theories}
\message{Decoupling Theories}

We will assume here that the high-energy theory is a gauge theory
consisting of scalars, fermions, and/or vectors.  We also assume that
high-energy phenomena decouple from the Standard Model, so that we may use
a linear effective Lagrangian with an explicit Higgs field.  We will divide
the effective operators into those which can be generated by tree graphs in
the underlying theory, and those which can be generated only through loop
diagrams.  A four-fermion operator, for example, can be produced at tree
level by any theory with a heavy gauge field coupling to Standard-Model
fermions.  A heavy fermion, though, produces four-fermion operators only at
the one-loop level (figure \fourfermi). Since we don't know the form of the
underlying theory, we will
take the broadest possible path: those operators which can be generated at
tree level by {\it some} underlying (gauge) theory will be separated from
those which can never be generated at tree level.


\vskip.2in
\box23
\vskip.1in


Buchm\"uller and Wyler [\bw] have assembled a list of all dimension-six
effective operators (assuming baryon and lepton conservation,
there are none at dimension five).  In assembling this
list, the authors have used the classical equations of motion to drop a
number of operators.  The use of the classical EOM is allowable; although
operators related by them give different Green's functions, they lead to
identical S-matrix elements [\canlag].  But the comparison we will make
here is between {\it Green's functions} in the effective theory and {\it
Green's functions} in the full theory, and so we must consider {\it all}
possible effective operators.  If we considered only the reduced set of
operators, we would ignore that fact that, in eliminating some operator
with the EOM, we also mix up the coupling constants of the various
operators [\canlag].  If we remove an operator with a large coupling
constant, this large effect will be transferred to other operators, and
we need to keep track of this.

\section{Possible Vertices}
\message{Possible Vertices}

Now we will attempt to estimate the size of the effective coupling
constants. The first step is to assemble a list of all possible tree graphs
in the full theory which contribute to dimension-five or dimension-six
effective operators.  These are graphs with only heavy internal lines and
only light external lines.  Power counting limits the number of such
graphs.  Each internal boson gives a factor $(p^2 - \Lambda^2)^{-1}
\rightarrow 1/\Lambda^2$ (for momenta $p << \Lambda$), and each internal
fermion gives a factor of $1/\Lambda$.  We use the most general set of
vertices
(figure \allvertices) up to dimension four.  It may be that the
theory valid at energies in the TeV range is itself an effective theory.
Even so, operators in the high-energy theory of dimension five and larger
will be suppressed by powers of a mass scale large compared to the already
large mass scale of the high-energy theory, so they can safely be
neglected. All relevant graphs are listed in figure \allgraphs; solid lines
represent fermions, dashed lines represent scalars or vectors.  Dotted
three-boson vertices must have a coupling constant on the order of
$\Lambda$, the heavy scale; undotted ones need not.

\vskip .3in
\box35 
\vskip .2in

We now apply the restriction that the full theory obey the constraints of
gauge invariance.  It simplifies matters to consider the $\su3_C \times
\su2_L \times \ui_Y$ symmetry of the low-energy effective Lagrangian to be
unbroken at this stage.  Only after matching the effective Lagrangian to
the high-energy theory and determining the strengths of the effective
couplings will we break $\su2_L \times \ui_Y \rightarrow \ui_{Q}$.  We will
find that this restriction prohibits certain vertices.

We will refer to the Standard Model fields as ``light" fields, and to the new
fields, with masses above some heavy scale $\Lambda$, as ``heavy".  $F_\mu^a$
will refer to all vectors, both light and heavy. $\pl$, $\fl$,  and $A_\mu^a$
will refer to light scalars, fermions, and vectors, respectively, and $\ph$,
$\fh$,  and $X_\mu^a$ will denote the heavy fields; note that these refer to
the {\it mass} eigenstates (before $SU(2)\otimes U(1)$ breaking.)
We consider the various types of vertices one by one:

\underbar{Vector interactions}$\ \ $In a gauge theory,
vector-boson interactions come from $F^a_{\mu\nu}F^{\mu\nu}_a$, where
$ F^a_{\mu\nu} = \partial_\mu F^a_\nu - \partial_\nu F^a_\mu - g
f_{abc} F^b_\mu F^c_\nu $ is the field strength, and the Lie
commutators $ [T_a,T_b] = i f_{abc} T_c $ define the structure
constants $ f_{abc} $.

The $T_i$'s are the generators of the gauged symmetry; the light (heavy)
vectors have generators $T_l$ ($T_h$).  Each vertex $F^a F^b F^c$ thus comes
with one derivative and a factor of $f_{abc}$.  Each vertex  $F^a F^b F^c F^d$
comes with a factor of $f_{abe} f_{cde}$.    The Standard-Model symmetry
$\su3_C \times \su2_L \times \ui_Y$, generated by the $T_l$, forms a subgroup
of the full symmetry, generated by $T_l + T_h$.  It follows that $[T_l,T_l]
\propto T_l$, and so $f_{llh} = 0$. $f_{hhl}$, $f_{lll}$, and $f_{hhh}$ may be
nonzero, and so the vertices XXA, AAA, and XXX are possible.  The non-zero
combinations $f_{abe} f_{cde}$ are: \setbox1= \vbox{
<LE>}
$$ \vbox {\settabs 2 \columns
\+ $f_{lll}f_{lll}$  & $f_{hhl}f_{hhl}$ \cr
\+ $f_{hhl}f_{lll}$  & $f_{hhh}f_{hhh}$ \cr
\+ $f_{lhh}f_{lhh}$  & $f_{lhh}f_{hhh}$, \cr }
\eqn\fourveq
$$
\setbox1= \vbox{
<LE>}
leading to the vertices AAAA, XXXX, XXAA, and AXXX.

\underbar{Scalar interactions}$\ \ $  The heavy scalars may or may not have a
non-zero VEV and may get masses from spontaneous symmetry breaking,
or from an explicit mass term, or both.  The light scalars, of course, are the
Standard Model $\su2_L$ doublets.

First examine the three-scalar terms.  A $\pl\pl\pl$ vertex is impossible
because it breaks $\su2_L$.  Allowing for the possibility of heavy $\su2_L$
singlets and doublets (with and without a VEV), $\ph\ph\ph$, $\pl\ph\ph$, and
$\pl\pl\ph$ vertices are all possible, each with a coupling proportional to a
light or heavy mass scale, depending on the underlying theory. All four-scalar
vertices are possible.

\underbar{Vector-Fermion interactions}$\ \ $ These enter only through the
covariant derivative terms $ \sum_i\psi_i\Dslash\psi_i  $ where $
D_\mu = \partial_\mu  + ig\thinspace T^a F^a_\mu. $

The $\psi_i$ are fermion multiplets, possibly containing both heavy and light
fermions.  All but one of the various combinations of light and heavy fermions
and vectors are in general possible.  $\fl\fh A$ does not occur, since $T_l$,
being unbroken, cannot mix light degrees of freedom with heavy ones.

\underbar{Scalar-Fermion interactions}$\ \ $The interaction term is \setbox1=
\vbox{
<LE>}
$$
\sum_{i,j,k}  y_{i,j,k} \psi_i\psi_j\phi_k.
\eqn\yukawa
$$
\setbox1= \vbox{
<LE>}
Any interaction of this form can be found in a general gauge-invariant field
theory.

\underbar{Scalar-Vector interactions}$\ \ $  These interactions come from the
term  \setbox1= \vbox{
<LE>}
$$
    \sum_i(D_\mu \phi_i)^\dagger (D^\mu \phi_i),
\eqn\scakin
$$
\setbox1= \vbox{
<LE>}
where $\phi_i$ is a scalar multiplet, possibly containing both heavy and light
scalars.  If $\phi_i$ is heavy and gets no VEV, or if it coresponds to the
Standad Model doublet, all terms of the form $ \phi \phi F F $ and
$ \partial \phi \phi F $ are allowed.
If $\phi_i$ contains both heavy and light scalars, then the
terms like $\partial\pl\ph X$ and $\pl\partial\ph X$ are also allowed.
In contrast $\partial\pl\ph A$ and
$\pl\partial\ph A$ are forbidden: $T_l$ generates the $\su2_L \times \ui_Y$
subgroup, and so it cannot mix heavy and light scalars.
Similarly the terms, $\pl\ph X X$ and $\pl\ph\ X A$ are allowed, but $\pl\ph\ A
A$ is
forbidden.

The vertices $\pl A A$, $\ph A A$, $\ph A X$, $\pl A X$, and $\pl X X$ are
forbidden.  These can only come from $(D_\mu \phi)^\dagger D^\mu \phi$
where $\phi$ gets a heavy VEV (i.e. some VEV giving mass to heavy
particles). The structure of this term is $\phi^\dagger T_a T_b v_h F^a_\mu
F^{b\mu}$, where F is a light or heavy vector, and $v_h =
\bra{0}\ph\ket{0}$.  Since the Standard-Model $\su2_L \times \ui_Y$ is
unbroken, $T_l v_h = 0$, thus eliminating $\pl A A$ and $\ph A A$.  We will
choose to work in the unitary gauge, and so, since $ \phi T_l $ is a
vector in the Goldstone direction, the couplings $\pl X A$ and $\ph X A$ do
not exist.  Finally, $\pl X X $ is absent, though the coupling $\ph X X$
may exist.

In summary, the forbidden vertices are:
\setbox1=
\vbox{ <LE>}
$$
\matrix{\pl\pl\pl       &\pl\ph A A  &\pl\ph A    &\pl A X   \cr
        \pl A A         &\ph A A     &\ph A X     &\pl X X   \cr
        AAX             &AAAX        &\fl\fh A    &          \cr}.
\eqn\forbidden
$$
\setbox1= \vbox{
<LE>}

\section{Tree-Level Effective Operators}
\message{Tree-Level Effective Operators}

Now that all the possible vertices have been enumerated, we can put
them together in all possible ways (figure \allgraphs), and produce
all possible tree-level low-energy effective operators.  We have
plenty of flexibility to create tree graphs giving contributions to
many of the various effective operators (all of which are listed in
Appendix A).  As previously noted, there are no dimension-five
operators assuming lepton- and baryon-number conservation; the
possibilities start only at dimension six.

\vskip .3in
\box16 

Care must be taken to perform this process in a consistent manner, giving
both ends of each propagator only one set of quantum numbers at a time.  It
is this restriction which ensures that no dimension five operators are
found (the graphs in figure \allgraphs\ (b), (i) and (k) cannot be created,
and (e) can only be created along with a $\gamma_\mu \partial^\mu$).
Dimension-four operators can be created in this way, but the gauge symmetry
insures that they only serve to renormalize the Standard-Model terms.

It is clear from the beginning that, since no graph in figure \allgraphs\
has only three legs, an operator which has any three-field part cannot come
at tree level (see Appendix C for a discussion).  By gauge invariance the
other parts of the operator, four-, five-, and six-field parts, must also
not come at tree level.  The operators excluded in this way are operators
(A.3) - (A.6), (A.30) - (A.37), and (A.57) - (A.59).  We will see that no
parts of these operators are generated at tree level.

The four-fermi operators (A.7) - (A.26) can easily be created from figure
\allgraphs (a).  The scalars-only operators (A.27) - (A.28) can be created
from figure \allgraphs (h), (j), and (l) - (o).  Similarly, all the
operators in (A.44) - (A.55) can be created using various structures in
figure \allgraphs.

All the rest in Buchm\"uller and Wyler's catalogue cannot be generated in
this manner, and so, {\it for any underlying gauge theory, they are
generated at no lower than the one-loop level}.  The reason they can't be
tree graphs is straightforward.

First consider only graphs and vertices without fermions.  Every such
vertex with exactly one heavy line has at least two light scalars. Every
such tree graph in our list has at least two such vertices, and so every
tree graph in our list must have at least four light scalars. Operators
(A.40) - (A.43) have no fermions, and only two scalars, and so these
operators must come from loops in the underlying theory.  Operators (A.3) -
(A.6) have no fermions and no scalars, and so they too must come from
loops.  Similarly, the other trilinear gauge vertices (eliminated in [\bw]
by the equations of motion) come only at loop level.

Operators with two fermions can only come from figure \allgraphs\ (c) -
(g).  Graphs (c), (d), and (g) each have at least one vertex without
fermions, and with exactly one heavy line.  They must therefore have at
least two external scalars.  Graphs (e) and (f) have two vertices which
contain a light fermion, a heavy fermion, and a light boson; this boson can
only be a scalar. The conclusion is that each of these graphs can only be
created with at least two external scalars.  The operators in (A.30) -
(A.37) have two fermions and no scalars, and so no graphs give tree-level
contributions to them.  The operators in (A.57) - (A.64) have two fermions
and only one scalar, so these too must come from loops in the underlying
theory.

In summary, operators with no fermions must have at least four scalars to
come at tree level, and operators with two fermions must have at least two
scalars to come at tree level.  The operators in [\bw] satisfying these
conditions are (A.7) - (A.26), (A.27) - (A.28), and (A.44) - (A.55); {\it
only these} may be created at tree level.

\section{Equations of motion}
\message{Equations of motion}

We have now examined the genesis of each operator in Appendix A, but this
is not enough.  Appendix A contains a ``complete set" of operators, but
many others have been removed by use of the equations of motion. To confirm
that the coefficients of the ``loop" operators are really suppressed by $1
/ 16\pi^2$, we have to check the related operators which have been removed
by the equations of motion.  If any of these came at tree level, its
unsuppressed coefficient will be added to the effective couplings of the
remaining operators. The coefficient of a loop-level operator can get a
tree-level contribution from the removed operator.  If any of the operators
(A.3) - (A.6), (A.30) - (A.43), or (A.57) - (A.64) are linked in this way
to a missing operator, then they will have tree-level coefficients,
contrary to our previous claims.

First consider operators with no fermions.  To get tree-level
contributions, these operators must have at least four scalars.  The other
two powers of momentum may come from additional scalars or derivatives.
The possible Lorentz and gauge-invariant operators are: \setbox1=\vbox{
<LE>}
$$
\eqalign {
   (\phi^\dagger \phi)^3 &= 3 \ocal_\phi    \cr
   \partial_\mu (\phi^\dagger \phi)
     \partial^\mu (\phi^\dagger \phi) &= 2\ocal_{\partial\phi} \cr
  (\phi^\dagger D_\mu \phi)
     (D^\mu \phi^\dagger\phi) &= \ocal_\phi^{(3)}   \cr
  (\phi^\dagger \phi)
     (D_\mu \phi^\dagger D^\mu \phi) &= \ocal_\phi^{(1)}   \cr
   (\phi^\dagger D_\mu \phi)
   (\phi^\dagger D^\mu \phi) &=
              -\ocal_{\phi}^{(3)} - \ocal_\phi^{(1)}
    - (\phi^\dagger \phi) (\phi^\dagger D^2\phi)
               \ \ \ \ \ (IBP)  \cr
    (\phi^\dagger \phi) (\phi^\dagger D^2\phi)  &=
       m^2(\phi^\dagger\phi)^2 - 3\lambda\ocal_\phi
                     -y_e \ocal_{e\phi}^\dagger
                     -y_u \ocal_{u\phi}^\dagger
                     -y_d \ocal_{d\phi}^\dagger \ \ \ \ \ \ (EOM)     \cr}
\eqn\EOMtwo
$$
\setbox1=\vbox{
<LE>} and their Hermitian conjugates.  The equation labeled IBP was
obtained by integration by parts, that labeled EOM using the equations of
motion.  Of all of these the only operator to be removed by the equations
of motion is the last one; it is converted into the operators on the right
hand side of the equation.  Each of these was already a tree-level
operator, and so no damage is done to our previous classifications (the
$\left(\phi^\dagger\phi\right)^2$ term just renormalizes the Standard
Model).

Operators with two fermions can only come at tree level if they have at
least 2 scalars.  Lorentz and gauge invariance imply that the other power
of momentum can only be a derivative or a scalar.  The possible operators
(not including Hermitian conjugates) are:
\setbox1=\vbox{
<LE>}
$$
\eqalign{
   (\bar l \gamma^\mu l) (\phi^\dagger D_\mu\phi)
       &= -i\ocal_{\phi l}^{(1)}\ \ \hbox{and similarly for}\
       \ocal_{\phi q}^{(1)}, \ocal_{\phi e}, \ocal_{\phi u}, \ocal_{\phi d}\cr
   (\bar l \gamma^\mu \tau^I l) (\phi^\dagger D_\mu \tau^I \phi)
       &= -i\ocal_{\phi l}^{(3)}\ \ \hbox{and similarly for}\
       \ocal_{\phi q}^{(3)}     \cr
   (\bar u \gamma^\mu d) (\phi^\dagger \epsilon D_\mu \phi)
                                  &= -i \ocal_{\phi\phi} \cr
   (\bar l e \phi) (\phi^\dagger\phi)
       &=  \ocal_{e\phi} \ \ \hbox{and similarly for}\
          \ocal_{u\phi}, \ocal_{d\phi}  \cr
   (\bar l \gamma^\mu l) \partial_\mu(\phi^\dagger\phi)
     &= - (\phi^\dagger\phi) (\Dslash\bar l l )
      - (\phi^\dagger\phi) (\bar l \Dslash l )
                         \ \ \ \ \ IBP \cr
   (\phi^\dagger\phi) (\bar l \Dslash l )
     &= -iy_e \ocal_{e\phi} \ \ \hbox{and similarly for}\
    \ocal_{u\phi}, \ocal_{d\phi}
        \ \ \ \ \ EOM \cr   }
\eqn\EOMthree
$$
\setbox1=\vbox{
<LE>} The operators in this set removed by the equations of motion are $
(\phi^\dagger\phi) (\bar f \Dslash f)$ for any fermion multiplet $f$.
These are traded in for $\ocal_{e\phi}$, $\ocal_{u\phi}$, and
$\ocal_{d\phi}$, which are tree-level operators in their own right, and so
again there is no change in the previous classification. All operators with
four fermions are already in the list [\bw], so we need not worry about
them.

The conclusion, then, is that for the choice of operators removed by the
equations of motion in [\bw], no loop-level operators get tree-level
contributions from the removed operators.  In this reduced set of effective
operators, (A.3) - (A.6), (A.30) - (A.43), and (A.57) - (A.64) are indeed
suppressed by small coefficients of order $1/16\pi^2$.  If a different set
of operators had been removed by the EOM, this conclusion might no longer
be true.  Of course, if ALL operators were included in the set, including
those related by the EOM, then this complication would not arise.

\chapter{Dimension-eight operators}
\message{dimension-eight operators}

After symmetry breaking, there are many vertices which cannot be generated
at tree level, for example the trilinear gauge vertices $W^+ W^- A$ and
$W^+ W^- Z$.  Nonetheless, any vertex allowed by the symmetry of the theory
can be generated at tree level if we allow for operators of high enough
dimension. For example, the dimension-eight operators
\setbox1= \vbox{
<LE>}
$$  \eqalign {
\ocal_{8,1} &= \left(\phi^\dagger \tau^I \phi \right)
          \left(\phi^\dagger \tau^J \phi \right)
                      W^I_{\mu\nu}W^{J\mu\nu}  \cr
\ocal_{8,2} &= \left(D^\mu \phi^\dagger D^\nu \phi \right)
          \left(\phi^\dagger \tau^I \phi \right)
                     W^I_{\mu\nu}  \cr
}
\eqn\deight
$$
\setbox1=\vbox{
<LE>}
contain trilinear gauge vertices. These may be generated at tree level; parts
of these operators come from the tree graphs in figure \dimeight.


\vskip.2in
\box43
\vskip.1in


These are suppressed (relative to the dimension six operators) by a factor
of $v^2/2\Lambda^2$, but may not be suppressed by a loop factor $ \sim
1/ 16 \pi^2 $.  If the
scale $\Lambda$ at which new physics enters is less than about
$2\sqrt{2}\pi v \sim 2
\tev$, {\it the effects of these dimension-eight operators will be competitive
with, or larger than, those of the dimension-six loop vertices}.  Thus
relationships like [\hagis] \setbox1=\vbox{
<LE>}
$$ \eqalign {
\delta\kappa_Z + {sin^2\theta_w \over cos^2\theta_w}\delta\kappa_A
                               & = \delta g_1^Z \cr
\lambda_A &= \lambda_Z \cr
}
\eqn\klgrel
$$
\setbox1=\vbox{
<LE>} which are true for dimension-six operators (but not for
dimension-eight operators) may not be reproduced by the data, even for
decoupling high-energy physics.

\chapter{Chiral Lagrangian}
\message{Chiral Lagrangian}

In the previous chapter we considered an effective Lagrangian built from
fields including a light Higgs doublet.  We may also consider a Lagrangian
containing only Standard-Model fields, not including a Higgs.  We know that
there are Goldstones in the low-energy spectrum, since we have seen massive
gauge bosons, and so any effective theory must include these states.  The
effective Lagrangian including all Standard-Model fields, including the
Goldstones but excluding the Higgs, is a chiral lagrangian.  This makes use
of a nonlinear representation for the Goldstones so as to include them, but
not the Higgs.  The resulting effective Lagrangian, including all terms
with the appropriate symmetry (but excluding fermion interactions) is just
that of Appendix B.

If the full physical theory includes a weakly-coupled Higgs boson, then we
may use either the linear effective Lagrangian containing it, or the chiral
Lagrangian without it. (Which one we use would depend on the spectrum of other
particles.)  In the event that the Higgs is heavy (i.e. at least as heavy
as other non-Standard-Model particles) then we must use the chiral
Lagrangian.  If the full theory is non-decoupling, strongly coupled, or if
it contains no fundamental Higgs scalar at all (as in technicolor models) then
we
are forced to use the chiral Lagrangian.

If we exclude operators containing fermions, then at lowest non-trivial
order the list of effective operators reduces to the twelve enumerated
in Appendix B.  Nine of these explicitly violate the custodial $SU(2)_R$;
the three $SU(2)_R$-violating operators which contain two-vector terms
($\lcal_1', \lcal_1,$ and $\lcal_8$) modify the relation $\rho= (M_W/M_Z
\cos\theta_w)^2 = 1$.  LEP measurements show $\rho -1 < 0.005$ [\alt], and
so, phenomenologically, $\beta_1', \beta_1, \beta_8 \lesim 1/16\pi^2$.

Of the remaining nine operators in $\lcal_{eff}$, five ($\lcal_4, \lcal_5,
\lcal_6, \lcal_7,$ and $\lcal_{10}$) contain only four-boson couplings.  To
find their coefficients, we must equate the Green's functions produced by
the effective theory with those from the full theory.  This is just as in
the last section, except that now we carry out this matching with the
low-energy symmetry $SU(2)_W \times U(1)_Y$ {\it broken}.  The four-point
operators can all be generated at tree level by some possible nondecoupling
theory, using the graph \allgraphs\ (h).  Four operators ($\lcal_2,
\lcal_3, \lcal_9, \lcal_{11}$) remain to be considered.

One example of perturbative nondecoupling physics is simply the
Standard Model, with a heavy (though still perturbative) Higgs Boson
which does not decouple from the other massive particles.  In this
case, integrating out the Higgs gives only $\beta_5$ at tree-level,
$\beta_1', \beta_1 - \beta_4 \propto 1/16\pi^2$, and all others zero
[\her].  Another typical scenario for perturbative nondecoupling
physics is a heavy fourth family.  In this case we could integrate out
the heavy fermions as well as the Higgs.  Any diagram with only light
vectors as external lines, and with at least one heavy fermionic
internal line, must be a loop diagram.  All the couplings will, as in
the previous example, be of order $ 1 / 16 \pi^2 $ or less, with the
exception of $\beta_5$.  In general a theory with heavy scalars can
give four-vector operators at tree level, but the only way to get the
three-point vertices in $\lcal_2, \lcal_3, \lcal_9,$ and $\lcal_{11}$
at tree level is through mixing between the light vectors and some
heavy vector.  This could come about only if the Higgs giving mass to
the light vectors also transforms non-trivially under another gauged
group.  The heavy gauge bosons corresponding to this new group must be
massless when $SU(2) \times U(1)$ is unbroken (they could be heavy
because of a relatively large coupling constant).  If they get their
heavy mass from symmetry breaking at a higher scale, they would be
decoupling, and the analysis of the previous section would apply.

If the physics beyond the Standard Model is strongly coupled it becomes
harder to make firm statements.  The dimension-two Goldstone terms in the
chiral Lagrangian have coupling constants equal to $v^2$.  The
dimension-four terms should be suppressed by two powers of the scale of
heavy physics, so $\beta_i \simeq v^2 / \Lambda^2$.  The unknown here is
$\Lambda$.  The usual method of estimation is ``Naive Dimensional
Analysis'' (NDA) [\nda,\gbook].  This amounts to the demand that the
tree-level diagrams from dimension-four operators be no smaller than the loop
diagrams from dimension-two operators, for a robust choice of
renormalization scale.  This leads to the requirement that $\Lambda^2
\lesim 16\pi^2 v^2$, and to the ``natural'' scale of $\Lambda^2 \sim
16\pi^2 v^2$.  If this simplistic argument is correct, then $\beta_i \sim
1/16\pi^2$.  There is nothing in principle to forbid a smaller $\Lambda$, but
we should
note that these predictions are quite consistent with low-energy QCD
phenomenology [\gla,\glb].  Of course it might also be said that this is a
coincidence; that the parameters of the QCD chiral Lagrangian can be fitted
by matching on to the lowest-lying resonances in the theory.  Then
vector-meson dominance, combined with the fact that the mass of the $\rho$
(the lowest-lying vector resonance) is about the same as $4 \pi f_\pi$,
gives the same answer.  So we need to ask if a resonance much below $4\pi
v$ is possible.

As a concrete example consider the dynamics of a strongly-coupled Higgs
sector.  Such a theory might look like the Standard Model, but with a large
Higgs' self-coupling $\lambda$.  As $M_H \propto \lambda v$ grows larger
than about $1 - 2\ \tev,$ perturbative unitarity breaks down at $\sqrt{s}
\sim 1 - 2\ \tev$ [\lee].  Since we know the theory is unitary for any
$\lambda$, this indicates the breakdown of perturbation theory; there must
be some new phenomenon which acts to unitarize the theory before such
energy scales are passed.  Presumably the effects of this new physics come
in around threshold energies of $1-2 \tev$.  $ 1/N $ calculations indicate
the existence of a single Higgs resonance at about $900 \gev$ as $\lambda
\rightarrow \infty$ [\einb].  There is only a single scalar resonance in
the theory.  Four-point Green's functions can occur at tree-level by
exchange of the scalar.  Three-point Green's functions could only come from
mass eigenstate mixing between Standard-Model gauge bosons and heavy
excitations.  But the heavy scalar excitation can't mix with the
Standard-Model vectors, and so three-point functions can't come at
tree-level.  The one tree-level parameter is $\beta_5 \sim v^2 / (8 M_H^2)$
[\gla,\bagger,\dona].  For $M_H = 900 \gev$, $\beta_5 \sim .009 \sim
1/16\pi^2$, in agreement with NDA.  The other parameters are of the same
order or even smaller.

Another example of a strongly-coupled symmetry-breaking sector is
technicolor.  Several estimates of the strength of the effective couplings
have been made, all of which give results consistent with $\beta_i \sim
0.001 - 0.05$.  NDA in this case gives $\beta_i \sim N_{TF} N_{TC}
/16\pi^2$, where $N_{TF}$ is the number of technifermions, and $N_{TC}$ is
the number of technicolors [\nda,\holter].  The spread in $\beta_i$
corresponds mostly to the assumed specifics of the technicolor model.
Quark-loop calculations give the same results [\apwu,\qlc].  We can also
estimate the strengths of effective couplings by requiring the low-energy
cross sections to match onto the tails of the lowest-lying resonances
[\dona].  This approach works quite well for QCD [\gla,\dona,\ecker].
Technicolor resonances typically appear at $\ocal(1 \tev)$, giving
estimates for $\beta_i$ in the range quoted above.  In this case, there are
vector resonances, such as the technirho, which may mix with the gauge
bosons, and so even the three-point operators $\lcal_2, \lcal_3, \lcal_9,$
and $\lcal_{11}$ can get contributions, at lowest order, from the
underlying physics.  Finally, we can use our knowledge of low-energy QCD to
give estimates for these parameters.  Assuming that, to first order,
technicolor is just scaled-up QCD, we can get estimates of $\beta_i$ by
letting $\Lambda_{QCD} \rightarrow \Lambda_{TC}$, and $f_\pi \rightarrow
v$, as well as allowing for differing numbers of (techni)colors and
(techni)families.  This again gives $\beta_i \sim 0.001 - 0.05$
[\holter,\pesta].  Even so, it does not seem impossible that a
lower-lying resonance exists, leading to significantly higher values of the
effective couplings.\foot{For example, a low-lowing vector resonance is
presumed to occur in the BESS model[\bess].}

In summary then, no definite limits may be placed on coefficients from a
completely general higher-energy theory.  But even in the case of
strongly-coupled or nondecoupling physics, estimates for the operators
($\beta_2, \beta_3, \beta_9, \beta_{11}$) containing three-point vertices
are consistently small, that is, in the range $1/16\pi^2\times (0.1 - 10)$.
These can serve as indications of what may be expected, though it must be
kept in mind that exceptions to these limits do exist.

\chapter{Trilinear Gauge Vertices}
\message{Trilinear Gauge Vertices}

\def\wwv{\scriptscriptstyle WWV}
\def\wwa{\scriptscriptstyle WWA}
\def\wwz{\scriptscriptstyle WWZ}
\def\ss#1{\scriptscriptstyle #1}

The self-couplings of gauge bosons have up until now been only very poorly
tested; at present the direct experimental tests only limit the couplings
to the same order of magnitude as predicted by the Standard Model
[\hagis,\current].  There has been much interest in tightening the
experimental bounds on these couplings, which will first be directly
measured at tree-level at LEP 200.  Nevertheless, there are convincing
reasons to believe that the Standard-Model $\su2_L \times \ui_Y$ is the
correct low-energy gauge symmetry [\einhorn,\der].  This symmetry fixes the
boson self-couplings, at least for the dimension four operators.  In that
case, the deviations we can hope to see are the result of
higher-dimensional effective operators.

The most general Lorentz-invariant CP-conserving Lagrangian which
pa\-ram\-e\-trizes the interactions of two W bosons with an A or Z boson is
[\hagp]:
\setbox1=\vbox{
<LE>}
$$ \eqalign {
{ \lcal_{\wwv} \over g_{\wwv} } &=
        i g_1^V \left( \hat W_{\mu\nu}^\dagger W^\mu V^\nu
                        - W_\mu^\dagger V_\nu \hat W^{\mu\nu} \right)
               + i\kappa_V W_\mu^\dagger W_\nu \hat V^{\mu\nu} \cr
&\qquad {i \lambda_V \over M_W^2}
          \hat W_{\lambda\mu}^\dagger \hat W_\nu^\mu \hat V^{\nu\lambda}
            + g_5^V \epsilon^{\mu\nu\rho\sigma}
               \left( W_\mu^\dagger
                   \rlap{$\overleftarrow \partial_{\kern -5pt \rho}$}
                     \hbox{$\overrightarrow \partial_{\kern -5pt \rho}$ \hss}
                      \kern -5pt W_\nu \right) V_\sigma \cr
}
\eqn\hagi
$$
\setbox1=\vbox{
<LE>}
where $V$ may be either $A$ or $Z$, $\hat W_{\mu\nu} = \partial_\mu W_\nu -
\partial_\nu W_\mu$, $g_{\wwa} = -e$, and $g_{\wwz} = -e\cot \theta_w$.  The
Standard-Model values are $g_1^A = g_1^Z = \kappa_A = \kappa_Z = 1,$ and $
\lambda_A = \lambda_Z = g_5^A = g_5^Z = 0$.  The relations $g_1^A = 1$ and
$g_5^A = 0$ are ensured by electromagnetic gauge invariance.

After substituting $\phi\rightarrow \phi + v/ \sqrt{2}$ (in the
decoupling case), or going to the unitary gauge (in the nondecoupling
case), various terms in the effective Lagrangians give the operators
in \hagi.  The operators containing explicit three-boson terms are
$\ocal_{W\phi\phi}$, $\ocal_{B\phi\phi}$, $\ocal_{WB}$, $\ocal_{\phi
W}$, $\ocal_{W}$, and $\ocal_{DW}$. In addition to the three-boson
terms, some of these operators contain two-boson terms. These must
also be considered; they may change the relationship between
$W^3,B,Z,$ and $A$, and so they affect the three-boson couplings.
Finally, some of these operators change the relationship between
measured input parameters and $g,g',$ and $v$.  We will take as our
input parameters $e^*, M_Z^*,$ and $G_F^*$, fixed respectively by the
zero-momentum coupling of the photon to the electron, LEP data, and
low-energy muon decay; the starred parameters are those actually
measured experimentally.  Let $e\equiv g \sin\theta_w \equiv gg' /(g^2
+ {g'}^2)$.  Then we have
\setbox1=\vbox{
<LE>}
$$ \eqalign {
   e^* & \equiv - gs \left[ 1 +  {v^2 \over \Lambda^2}
                \left(       {c^2 \over 2} \alpha_{\ss \phi B}
                               + {s^2 \over 2} \alpha_{\ss \phi W}
                               - sc \alpha_{\ss WB}
                          \right)\right]
                        = -.313  \cr
   M_Z^* & \equiv {g^2 + {g'}^2 \over 4 v^2}
                \left[ 1 +    {v^2 \over \Lambda^2}
                  \left(
                        {c^2 \over 2} \alpha_{\phi W}
                      + {s^2 \over 2} \alpha_{\phi B}
                      +  sc           \alpha_{WB}
                      + {g^2 \over 4} \alpha_{DW}
                                                        \right)\right]
                                                        = 91.2 \gev  \cr
 G_F^* & \equiv  { 1 \over \sqrt{2} v^2}
                         = 1.17 \times 10^{-5} \gev^{-2}   \cr
}
\eqn\inputs
$$
\setbox1=\vbox{
<LE>} with $ s\equiv \sin\theta_w \equiv g'/ \sqrt{g^2 + g'^2}$ and $
c\equiv\cos\theta_w \equiv g/ \sqrt{g^2 + g'^2}.$ We define
$\kappa_V,\lambda_V, g_1^V,$ and $g_5^V$ to be the coefficients of the
appropriate tensors in $\lcal_{eff}$, divided by $g_{\wwv}^*$:
\setbox1=\vbox{
<LE>}
$$ \eqalign {
      g_{\wwa}^* & \equiv - e^*  \cr
     g_{\wwz}^* & \equiv - e^* \cot\theta_w^* \equiv
             - e\cot\theta_w \left[ 1  +  {v^2 \over \Lambda^2}
               \left(       - {s^2 \over 2} \alpha_{\ss \phi B}
                            + {2 - c^2 \over 2} \alpha_{\ss \phi W}
                            + {2s^3 c + sc \over c^2 - s^2} \alpha_{\ss WB}
                                         \right)  \right]. \cr
}
\eqn\gwwv
$$
\setbox1=\vbox{
<LE>} With these definitions the decoupling effective Lagrangian gives
\setbox1=\vbox{
<LE>}
$$ \eqalign {
\Delta g_1^A \equiv g_1^A -1 &= 0   \cr
\Delta g_1^Z \equiv g_1^Z -1 &=   {v^2 \over \Lambda^2}
            \left[ { s \over c(s^2-c^2) } \alpha_{WB}
                  + {g \over 4 c^2} \alpha_{W\phi\phi} \right] \cr
\Delta\kappa_A \equiv \kappa_A - 1 &=  {v^2 \over \Lambda^2}
            \left[ { g \over g'} \alpha_{WB}
                  + {g^2 \over 4g'} \alpha_{B\phi\phi}
                  + {g \over 4} \alpha_{W\phi\phi} \right] \cr
\Delta\kappa_Z \equiv \kappa_Z - 1 &=  {v^2 \over \Lambda^2}
            \left[ { 2sc \over s^2-c^2} \alpha_{WB}
                  - {g' \over 4} \alpha_{B\phi\phi}
                  + {g \over 4} \alpha_{W\phi\phi}  \right] \cr
\lambda_A = \lambda_Z &=  {v^2 \over \Lambda^2}
                \left[ {3g \over 2} \alpha_W
                        +  {3 g^2 \over 4}  \alpha_{DW} \right].  \cr
g_5^A = g_5^Z &= 0. \cr
}
\eqn\transdec
$$
\setbox1=\vbox{
<LE>}
Similarly, for the nonlinear operators [\apwu,\holdom]: \setbox1=\vbox{
<LE>}
$$ \eqalignno {
\Delta g_1^A \equiv g_1^A -1 &= 0   \cr
\Delta g_1^Z \equiv g_1^Z -1 &=
            \left[ {e^2 \over s^2(c^2-s^2)} \beta_1'
                  + {e^2 \over c^2(c^2-s^2)} \beta_1
                  + {e^2 \over s^2 c^2} \beta_3 \right] \cr
\Delta\kappa_A \equiv \kappa_A - 1 &=  { e^2 \over s^2 }
            \left[ -\beta_1 + \beta_2 + \beta_3 -\beta_8 +\beta_9 \right] \cr
   \Delta\kappa_Z \equiv \kappa_Z - 1 &=
            \left[ {e^2 \over s^2(c^2-s^2)} \beta_1'
                  + {e^2 \over c^2(c^2-s^2)} \beta_1
                  + {e^2 \over c^2} (\beta_1 - \beta_2)
                  + {e^2 \over s^2} (\beta_3 - \beta_8 + \beta_9) \right] \cr
\lambda_A = \lambda_Z &=  0  \cr
g_5^A &= 0  &(4.5)\cr
g_5^Z &=  { e^2 \over s^2 c^2 } \beta_{11}. \cr
}
$$
\setbox1=\vbox{
<LE>}

Precision measurements have up until now provided bounds of order 1 on the
parameters in \hagi, though some of the individual effective couplings
$\alpha$ and $\beta$ have been constrained a hundred times more stringently
[\der,\hagis,\current]. Higher energy experiments are expected to provide
more stringent bounds.  LEP 200 is expected to set $\delta\kappa_V,
\lambda_V \lesim 0.1$ [\barb ,\bilenky,\kane].  Similarly, NLC 500 can
provide upper limits around $0.03$ [\barklow,\burke], NLC 1000 can give
limits of about $0.005$ [\barklow], and LHC and SSC can give limits of
about $0.01$ [\bagger,\kane,\falk].

We have seen in the last two sections that the expected value of each
$\alpha$ and $\beta$ appearing in these equations is $ 1/16\pi^2$.
Furthermore, with each gauge boson in an effective operator should come a
factor of g or g', since we have assumed the underlying theory to be a
gauge theory.  So we should expect
\setbox1=\vbox{
<LE>}
$$
\matrix{
    {\displaystyle\alpha_{\scriptscriptstyle DW}} \sim {g^2 \over 16\pi^2}
     &\displaystyle\alpha_{\scriptscriptstyle W} \sim {g^3 \over 16\pi^2} \cr
    \displaystyle\alpha_{\scriptscriptstyle W\phi\phi} \sim {g \over 16\pi^2}
 &\displaystyle\alpha_{\scriptscriptstyle B\phi\phi} \sim {g' \over 16\pi^2}\cr
 \displaystyle\alpha_{\scriptscriptstyle WB} \sim {gg' \over 16\pi^2}. & \cr
}
\eqn\tgvvalues
$$
\setbox1=\vbox{
<LE>} In the decoupling case, for $\Lambda = v = 240\ \gev$, we would
estimate the parameters in \hagi\ to be about $0.003$.  Even assuming such
a low scale for new physics, this is probably too small to be seen at any
of the colliders. But this estimate neglects the dimension-eight operators,
which, as we have noted in section (2.4), may be more important then those
of dimension-six.  These operators should give values of $\Delta\kappa,
\lambda,$ and $\Delta g$ larger than the dimension-six ones by a factor of
about $8\pi^2v^2 / \Lambda^2$.  For $\Lambda = 240\ \gev$, this implies
trilinear gauge couplings of about $0.2$, though the effective Lagrangian
is not convergent for such small values of $\Lambda$.  Nevertheless, this
is within the range of sensitivity at each of the machines considered
above.  If we expect the latter three machines to directly observe any
particles of $500\ \gev$ or less, then $\Lambda$ should be at least $500\
\gev$ and so the estimate for $\Delta\kappa,\lambda,\Delta g$ drops to
about $0.01$.  This is on the edge of observability at SSC, LHC, and NLC
1000.  Stated another way, the reach of these experiments is only about
$300\ \gev$ at LEP 200, $400\ \gev$ at NLC 500, $500\ \gev$ at SSC and LHC,
and $600\ \gev$ at NLC 1000. None of these machines can see the effects on
the trilinear gauge vertices of physics much beyond the energies of the
machines themselves.

\vskip .3in


\setbox40 = \vbox {
\vbox{\tabskip=0pt \offinterlineskip
\hrule
\halign to 6.5in{& \tabskip=0pt\vrule\hskip1pt\vrule# &\tabskip=1em
plus5em minus1em # & \strut\hfil#\vphantom{${y_y}_y$\vrule height13pt}\hfil\cr
%
%
&&\batchmode\vbox to 0pt{\vskip8pt\hbox{Collider}}\errorstopmode
&&& \batchmode\vbox to 0pt{\vskip4pt\hbox{Sensitivity}\vskip.5pt
   \hbox{to $\kappa,\lambda,g$ }}\errorstopmode &&&
\multispan{10} Expected Value of $\kappa,\lambda,g$ & \cr
height1pt && \omit &&& \omit && \multispan{11} \kern-3pt \vrule height.4pt
width299pt & \cr
height2pt && \omit &&& \omit && \multispan{11} \kern-3pt \vrule height.4pt
width299pt & \cr
height5pt&&\omit&&&\omit&&&\omit&\omit&&\omit&&&\omit&\omit&&\omit&\cr
&&\omit &&& \omit &&& \multispan4 Decoupling $\lcal_{eff}$ &&& \multispan 4
Chiral $\lcal_{eff}$& \cr
height1pt&&\omit&&&\omit&&&\omit&\omit&&\omit&&&\omit&\omit&&\omit&\cr
&&\omit &&& \omit &&& $\Lambda = 500\ \gev$  &\omit&&  $\Lambda = 1\ \tev$
&&& $N = 30$ &\omit&& $N=3$ & \cr
height1pt&&\omit&&&\omit&&&\omit&\omit&&\omit&&&\omit&\omit&&\omit&\cr
\noalign{\hrule}
height2pt&&\omit&&&\omit&&&\omit&height-20pt&&\omit&&&\omit&height-20pt&&\omit&\cr
\noalign{\hrule}
height2pt&&\omit&&&\omit&&&\omit&height-20pt&&\omit&&&\omit&height-20pt&&\omit&\cr
&& Present &&& 1 &&& 0.01 &height-20pt&& 0.0007 &&& 0.05 &height-20pt&& 0.006 &
\cr
\noalign{\hrule}
&& LEP 200 &&& 0.1 &&& 0.01 &height-20pt&& 0.0007&&& 0.05 &height-20pt&& 0.006
& \cr
\noalign{\hrule}
&& NLC 500 &&& 0.03 &&& 0.01 &height-20pt&& 0.0007&&& 0.05 &height-20pt&& 0.006
& \cr
\noalign{\hrule}
&& SSC \& LHC &&& 0.01 &&& 0.01 &height-20pt&& 0.0007&&& 0.05 &height-20pt&&
0.006 & \cr
\noalign{\hrule}
&& NLC 1000 &&& 0.005 &&& 0.01 &height-20pt&& 0.0007&&& 0.05 &height-20pt&&
0.006 & \cr}
\hrule}
\setbox1=\vbox{
<LE>}
\vskip .1in \vbox{\singlespace \twelverm Table~1. \ninerm Sensitivites to new
trilinear gauge vertex couplings (see \hagi) are listed.  Also listed are the
expected values of these couplings, assuming dimension-eight tree-level
coefficients for the decoupling case.  Even for $\scriptstyle\Lambda = 500\
\gev$ (decoupling) or large N (nondecoupling), the couplings are just on the
edge of observability for LEP 500, SSC, and LHC.}}

\box40 
\vskip .2in

If there is new strongly coupled physics, we expect it to manifest itself at
scales of $\ocal (1\ \tev)$.  Given our estimates of $\beta_i  \sim
1/16\pi^2 \sim 0.006$, we can expect values of $\Delta\kappa_V,\lambda_V
\sim 0.006$.  This might be marginally detectable at NLC 1000, but is
too small to be seen elsewhere.  These numbers may be enhanced by factors
of N (particles or colors); an enhancement of ten would put them in range
of NLC 500, SSC, and LHC, and well above the sensitivity limit of NLC 1000.

Because deviations from the standard trilinear gauge boson couplings are
expected to be small, they are unlikely to be observed except perhaps at
NLC 1000 [\einhorn], or unless new decoupling physics exists at energies
just above those which can be directly observed.  Even a low scale of
decoupling physics, or a large enhancement in the nondecoupling case, might
only just be detected at NLC 500, SSC, and LHC; we should be unsurprised to
see no hint of anomalous vector boson couplings at these machines.  On the
other hand, we cannot rule out the possibility that types of nondecoupling
physics not considered may give larger effects.


\chapter{Summary}
\message{Summary}

Heavy physics which decouples from the Standard Model leaves traces of
itself through higher-dimensional effective operators which must be added
to those in the Standard Model.  We have examined all of the dimension-six
operators, under the assumption that the full theory is a gauge theory, and
estimated the sizes of their effective coefficients.  Some Green's
functions are produced by the full theory through tree diagrams; the
effective operators which reproduce these Green's functions must therefore
have large, $\ocal(1)$, coupling constants.  Some Green's functions are
created by the full theory only through loops; the effective operators
which mirror them must therefore have small, $\ocal(1 /16\pi^2)$ coupling
constants.

Having chosen a full basis of dimension-six effective operators (reduced by
using the equations of motion - see Appendix A), we have divided the
decoupling operators into those which may have tree-level coefficients (for
any possible high-energy theory), and those which can only be produced
through loops, regardless of the specifics of the high-energy theory.  We
find that the tree-level operators are (A.7) - (A.26), (A.27) - (A.28), and
(A.44) - (A.55).  All others have effective couplings suppressed by powers
of $1/ 16\pi^2$ and will therefore be negligible in comparison.  Of course,
for any given theory, only some of the possible ``tree-level'' operators
may actually be produced at tree level.

If we allow for more general forms of high-energy physics, by using a
chiral effective Lagrangian, then comprehensive statements are more
difficult to make.  We concentrate on operators not involving fermions
(Appendix B).  Here the lowest-order relevant operators are
chiral-dimension two and four.  There are twelve such operators (beyond
those replicating the Standard Model).  Of those, three ($\lcal_1',
\lcal_1,$ and $\lcal_8$) contribute to $\rho=(M_W/M_Z\cos\theta_w)^2$ and
can be experimentally limited to less than about $1 / 16\pi^2$.  Five
operators ($\lcal_4 - \lcal_7$ and $\lcal_{10}$) contribute only to
four-boson Green's functions, and may be created at tree-level.  The four
others, $\lcal_2, \lcal_3, \lcal_9,$ and $\lcal_{11}$, contain three-boson
terms; these must be more closely considered.  ``Naive Dimensional
Analysis'' (NDA) [\gbook,\nda] predicts coupling constants of about $v^2/
\Lambda^2 \sim N / 16\pi^2$ (where N is a multiplicity factor representing
the number of particles, or colors, contributing to the operators), and
several model-specific calculations give similar results.  The indications,
then, is for these couplings to be small ($\sim$ N/16$\pi^2$), though there
are certainly exceptions.

These results clearly have implications for precision measurements.  The
virtual effects of high-energy physics are usually broadly distributed.
The search for them should concentrate in areas where their effects may be
large and avoid those processes in which we expect the effects to be
limited.

Much attention has been directed at the triple-vector-boson vertices.
Dimension-six operators with triple-vector-boson couplings are produced at
loop level, while some of those of dimension-eight may produced at tree
level.  Contrary to expectations, then, the higher-dimen\-sion\-al
operators may have the more important effects (up to scales $\Lambda\sim 2\
\tev$).  Because the dimension-eight operators are suppressed by powers of
$(v/\Lambda)^4$, and the nondecoupling effects are likely to be reasonably
small, it is rather difficult to see their experimental signatures.  In the
decoupling case, for $\Lambda = 500\ \gev$ we expect $\Delta\kappa, \Delta
g$, and $\lambda$ to be about $0.01$ - undetectable at LEP 200, marginally
observable at NLC 500\foot{The Next Linear Collider at 500~GeV center-of-mass
energy.}
or LHC, and probably detectable at NLC 1000.
Because we are working with dimension-eight operators, the effects drop
rapidly with energy and for $\Lambda = 1\ \tev$ the trilinear gauge boson
parameters are about $0.0007$, beyond the range of any machine considered.
In the nondecoupling case the estimate is about $0.006$ for $\Delta\kappa$
and $\Delta g$, at the limit of sensitivity of NLC 1000 and too small to be
seen elsewhere.  Even if enhanced by large factors of $N = 30$, the effects
are only barely observable at NLC 500 or LHC, and still beyond the
range of LEP 200.  It may be that types of nondecoupling physics not
considered, or theories with large numbers of particles or colors, can give
effects larger than those we have considered here.  But the likeliest cases
seem to make prospects gloomy at LEP 200, and uncertain at NLC 500, SSC,
and LHC.

\ack
We wish to thank M. Herrero for pointing out some errors in our original
manuscript with regard to the results of Ref.~\her.
This work was supported in part by the U.S. Department of Energy (contracts
DE-FG03-92ER40701 for C. Arzt,  DE-AC02-76ER01112 for M.B.E., and
DE-AT03-76SF00010 for J.W.) and by the Texas National Research Laboratory
Commission.

\Appendix{A}
\titlestyle{\bf Linear Operators}
\vskip .2truein

The operators which follow comprise a complete set of operators for an
$\su3_C \times \su2_L \times \ui_Y$ invariant effective Lagrangian (the
hermitian conjugates of each operator are assumed present when necessary).
As many operators as possible have been removed by using the classical
equations of motion. This list was compiled by W. Buchm\"uller and D. Wyler
[\bw] and is reproduced here for completeness; the numbering system follows
that of [\bw].

\countdef\opnumber=20
\def\op #1 - #2 {\ocal_{#1} &= #2 &(A.\the\opnumber)\cr
               \global\advance\opnumber by 1}
\def\opa #1 - #2 - #3 {\ocal_{#1}^{#2} &= #3 &(A.\the\opnumber)\cr
               \global\advance\opnumber by 1}
\def\opc #1 - #2 - #3 - #4 {\ocal_{#1} &= #2 \qquad , \qquad \ocal_{#3} = #4
           &(A.\the\opnumber)\cr \global\advance\opnumber by 1}

\def\opd #1 - #2 - #3 - #4 - #5 - #6 {\ocal_{#1}^{#2} &= #3 \qquad , \qquad
                   \ocal_{#4}^{#5} = #6
                    &(A.\the\opnumber)\cr \global\advance\opnumber by 1}

\vskip .5in
\noindent{\ninerm A.1 VECTORS ONLY}
\global\opnumber=3
$$
\setbox1=\vbox{
<LE>}
\eqalignno{
\op G - {f_{ABC} G_\mu^{A\nu} G_\nu^{B\lambda} G_\lambda^{C\mu}}
\op {\tilde G} - {f_{ABC} \tilde G_\mu^{A\nu} G_\nu^{B\lambda}
G_\lambda^{C\mu}}
\op W - {\epsilon_{IJK} W_\mu^{I\nu} W_\nu^{J\lambda} W_\lambda^{K\mu}}
\op {\tilde W} - {\epsilon_{IJK} \tilde W_\mu^{I\nu} W_\nu^{J\lambda}
W_\lambda^{K\mu}}
}
$$
\setbox1=\vbox{
<LE>}
\noindent{\ninerm A.2 FERMIONS ONLY}
\setbox1=\vbox{
<LE>}
$$ \eqalignno{
\opd ll - (1) - {\half (\bar l \gamma_\mu l) (\bar l \gamma^\mu l)} - ll - (3)
    - {\half (\bar l \gamma_\mu \tau^I l) (\bar l \gamma^\mu \tau^I l)}
\opd qq - (1,1) - {\half (\bar q \gamma_\mu q) (\bar q \gamma^\mu q)} - qq
    - (8,1) - {\half (\bar q \gamma_\mu \lambda^A q) (\bar q \gamma^\mu
     \lambda^A q)}
\opd qq - (1,3) - {\half (\bar q \gamma_\mu \tau^I q) (\bar q \gamma^\mu \tau^I
    q)} - qq - (8,3) - {\half (\bar q \gamma_\mu \lambda^A \tau^I q) (\bar q
    \gamma^\mu \lambda^A \tau^I q)}
\opd lq - (1) - {\half (\bar l \gamma_\mu l) (\bar q \gamma^\mu q)} - lq - (3)
     - {\half (\bar l \gamma_\mu \tau^I l) (\bar q \gamma^\mu \tau^I q)}
\op ee - {\half (\bar e \gamma_\mu e) (\bar e \gamma^\mu e)}
\opd uu - (1) - {\half (\bar u \gamma_\mu u) (\bar u \gamma^\mu u)} - uu - (8)
     - {\half (\bar u \gamma_\mu \lambda^A u) (\bar u \gamma^\mu \lambda^A u)}
\opd dd - (1) - {\half (\bar d \gamma_\mu d) (\bar d \gamma^\mu d)} - dd - (8)
     - {\half (\bar d \gamma_\mu \lambda^A d) (\bar d \gamma^\mu \lambda^A d)}
\op eu - {\half (\bar e \gamma_\mu e) (\bar u \gamma^\mu u)}
\op ed - {\half (\bar e \gamma_\mu e) (\bar d \gamma^\mu d)}
\opd ud - (1) - {\half (\bar u \gamma_\mu u) (\bar d \gamma^\mu d)} - ud - (8)
   - {\half (\bar u \gamma_\mu \lambda^A u) (\bar d \gamma^\mu \lambda^A d)}
\op le - {(\bar l e)(\bar e l)}
\op lu - {(\bar l u)(\bar u l)}
\op ld - {(\bar l d)(\bar d l)}
\op qe - {(\bar q e)(\bar e q)}
\opd qu - (1) - {(\bar q u)(\bar u q)} - qu - (8)
        - {(\bar q \lambda^A u)(\bar u \lambda^A q)}
\opd qd - (1) - {(\bar q d)(\bar d q)} - qd - (8)
      - {(\bar q \lambda^A d)(\bar d \lambda^A q)}
\op qde - {(\bar l e)(\bar d q)}
\opa qq - (1) - {(\bar q  u)(\bar q d)}
\opa qq - (8) - {(\bar q  \lambda^A u)(\bar q \lambda^A d)}
\opc lq - {(\bar l e) \epsilon (\bar q u)}
        - {lq'} - {(\bar l u) \epsilon (\bar q e)}
}
$$
\setbox1=\vbox{
<LE>}
\noindent{\ninerm A.3 SCALARS ONLY}
\setbox1=\vbox{
<LE>}
$$ \eqalignno {
\op {\phi} - { {1 \over 3} (\phi^\dagger\phi)^3}
\op {\partial\phi} - {\half \partial_\mu(\phi^\dagger\phi)
             \partial^\mu(\phi^\dagger\phi)}
}
$$
\setbox1=\vbox{
<LE>}
\noindent{\ninerm A.4 FERMIONS AND VECTORS}
\global\advance\opnumber by 1
\setbox1=\vbox{
<LE>}
$$ \eqalignno{
\opc lW - {i \bar l \tau^I \gamma_\mu D_\nu l W^{I\mu\nu}}
            - lB - {i \bar l \gamma_\mu D_\nu l B^{\mu\nu}}
\op eB - {i \bar e \gamma_\mu D_\nu e B^{\mu\nu}}
\op qG - {i \bar q \lambda^A \gamma_\mu D_\nu q G^{A\mu\nu}}
\opc qW - {i \bar q \tau^I \gamma_\mu D_\nu q W^{I\mu\nu}}
            - qB - {i \bar q \gamma_\mu D_\nu q B^{\mu\nu}}
\op uG - {i \bar u \lambda^A \gamma_\mu D_\nu u G^{A\mu\nu}}
\op uB - {i \bar u \gamma_\mu D_\nu u B^{\mu\nu}}
\op dG - {i \bar d \lambda^A \gamma_\mu D_\nu d G^{A\mu\nu}}
\op dB - {i \bar d \gamma_\mu D_\nu d B^{\mu\nu}}
}
$$
\setbox1=\vbox{
<LE>}
\noindent{\ninerm A.5 SCALARS AND VECTORS}
\global\advance\opnumber by 2
\setbox1=\vbox{
<LE>}
$$ \eqalignno{
\opc {\phi G} - { \half (\phi^\dagger \phi)  G_{\mu\nu}^A G^{A\mu\nu} }
      - {\phi \tilde G} - { \half (\phi^\dagger \phi)  \tilde G_{\mu\nu}^A
              G^{A\mu\nu} }
\opc {\phi W} - { \half (\phi^\dagger \phi)  W_{\mu\nu}^I W^{I\mu\nu} }
      - {\phi \tilde W} - { \half (\phi^\dagger \phi)  \tilde W_{\mu\nu}^I
              W^{I\mu\nu} }
\opc {\phi B} - { \half (\phi^\dagger \phi)  B_{\mu\nu} B^{\mu\nu} }
      - {\phi \tilde B} - { \half (\phi^\dagger \phi)  \tilde B_{\mu\nu}
              B^{\mu\nu} }
\opc {WB} - { (\phi^\dagger \tau^I \phi)  W_{\mu\nu}^I B^{\mu\nu} }
      - {\tilde W B} - {(\phi^\dagger \tau^I \phi)  \tilde W_{\mu\nu}^I
              B^{\mu\nu} }
\opd {\phi} - (1) - {(\phi^\dagger \phi) (D_\mu\phi^\dagger D^\mu \phi)}
        - {\phi} - (3) - {(\phi^\dagger D^\mu \phi) (D_\mu \phi^\dagger \phi)}
}
$$
\setbox1=\vbox{
<LE>}
\noindent{\ninerm A.6 FERMIONS AND SCALARS}
\setbox1=\vbox{
<LE>}
$$ \eqalignno{
\op {e\phi} - {(\phi^\dagger \phi) (\bar l e\phi) }
\op {u\phi} - {(\phi^\dagger \phi) (\bar q u\tilde\phi) }
\op {d\phi} - {(\phi^\dagger \phi) (\bar q d\phi) }
}
$$
\setbox1=\vbox{
<LE>}
\noindent{\ninerm A.7 VECTORS, FERMIONS, AND SCALARS}
\setbox1=\vbox{
<LE>}
$$ \eqalignno{
\opa {\phi l} - (1) - {i (\phi^\dagger D_\mu \phi) (\bar l \gamma^\mu l)}
\opa {\phi l} - (3) - {i (\phi^\dagger \tau^I D_\mu \phi)
                          (\bar l \gamma^\mu \tau^I l)}
\op {\phi e} - {i (\phi^\dagger D_\mu \phi) (\bar e \gamma^\mu e)}
\opa {\phi q} - (1) - {i (\phi^\dagger D_\mu \phi) (\bar q \gamma^\mu q)}
\opa {\phi q} - (3) - {i (\phi^\dagger \tau^I D_\mu \phi)
                          (\bar q \gamma^\mu \tau^I q)}
\op {\phi u} - {i (\phi^\dagger D_\mu \phi) (\bar u \gamma^\mu u)}
\op {\phi d} - {i (\phi^\dagger D_\mu \phi) (\bar d \gamma^\mu d)}
\op {\phi \phi} - {i (\phi^\dagger \epsilon D_\mu \phi) (\bar u \gamma^\mu d)}
\global\advance\opnumber by 1
\opc De - {(\bar l D_\mu e) D^\mu \phi} - {\bar D e}
                             - {(D_\mu \bar l e) D^\mu \phi}
\opc Du - {(\bar q D_\mu u) D^\mu \tilde \phi} - {\bar D u}
                             - {(D_\mu \bar q u) D^\mu \tilde \phi}
\opc Dd - {(\bar q D_\mu d) D^\mu \phi} - {\bar D d}
                             - {(D_\mu \bar q d) D^\mu \phi}
\opc eW - {(\bar l \sigma^{\mu\nu} \tau^I e) \phi W_{\mu\nu}^I}
           - eB - {(\bar l \sigma^{\mu\nu} e) \phi B_{\mu\nu}}
\op uG - {(\bar q \sigma^{\mu\nu} \lambda^A u) \tilde \phi G_{\mu\nu}^A}
\opc uW - {(\bar q \sigma^{\mu\nu} \tau^I u) \tilde \phi W_{\mu\nu}^I}
           - uB - {(\bar q \sigma^{\mu\nu} u) \tilde \phi B_{\mu\nu}}
\op dG - {(\bar q \sigma^{\mu\nu} \lambda^A d) \phi G_{\mu\nu}^A}
\opc dW - {(\bar q \sigma^{\mu\nu} \tau^I d) \phi W_{\mu\nu}^I}
           - dB - {(\bar q \sigma^{\mu\nu} d) \phi B_{\mu\nu}}
}
$$
\setbox1=\vbox{
<LE>}
The trilinear gauge vertices removed from this list by the equations of motion
are
\setbox1=\vbox{
<LE>}
$$ \eqalignno{
\ocal_{B\phi\phi} &= i (D_\mu \phi)^\dagger (D_\nu \phi) B^{\mu\nu} &(A.65) \cr
\ocal_{W\phi\phi} &= i (D_\mu \phi)^\dagger \tau^I
                        (D_\nu \phi) W^{I\mu\nu}  &(A.66) \cr
\ocal_{DW} &= (D_\alpha W^{I\mu\nu})^\dagger (D^\alpha W_{\mu\nu}^I)  &(A.67)
\cr
\ocal_{\partial B} &= (\partial_\alpha B^{\mu\nu})
                        (\partial^\alpha B_{\mu\nu}). &(A.68)   \cr
}
$$
\setbox1=\vbox{
<LE>}

\Appendix{B}
\titlestyle{\bf Nonlinear Operators}
\message{Nonlinear Operators}
\vskip .2truein

The following is a list of operators for use in the nondecoupling case.   We
will assume here that the particle spectrum is the same as the Standard Model's
with the exception of the Higgs, so that the effective Lagrangian can be
written as a gauged chiral model [\ndc-\glb,\apwu,\long].

Specifically, if  \setbox1=\vbox{
<LE>}
$$
U=\exp \left[ 2i \pi^a \tau^a / v \right]
\eqn\eq
$$
\setbox1=\vbox{
<LE>}
then the lowest order kinetic terms in the effective lagrangian are
\setbox1=\vbox{
<LE>}
$$
\lcal \lowti{ kin } = { v^2 \over 4 } \tr \{ \DD_\mu U^\dagger \DD_\mu U \}
          - \half  \tr \{ \WW_{ \mu \nu } \WW^{ \mu \nu } \}
          - \half \tr \{ \BB_{ \mu \nu } \BB^{ \mu \nu } \}
\eqn\lkin
$$
\setbox1=\vbox{
<LE>}
where we have adopted the matrix notation $ \WW_\mu = W^I_\mu \tau^I / 2 $,
$\BB_\mu = B_\mu \tau_3 / 2 $, $ \DD_\mu U = \partial_\mu U + i g \WW_\mu U - i
g' U \BB_\mu$, and $ \tau^I $ denote the Pauli matrices.

There are twelve new CP-invariant $ \su 2_L \times \ui_Y $ operators which are
of chiral dimension four or lower [\long,\apwu]. The only term of chiral
dimension
two is
\setbox1=\vbox{
<LE>}
$$
\lcal_1'= {v^2 \over 4} \beta_1' \left( \tr \left[ \tau^3 U^\dagger
\DD_{\mu} U \right] \right)^2
\eqn\eq
$$
\setbox1=\vbox{
<LE>}
and the eleven of order [mass]$^4$ are
\setbox1=\vbox{
<LE>}
$$ \eqalign {
\lcal_1 &= g g' \beta_1 \tr \left[ U \BB_{ \mu \nu } U ^\dagger
\WW^{ \mu \nu } \right] \cr
\lcal_2 &= - 2 i g' \beta_2 \tr \left[ \BB_{ \mu
\nu } \DD^\mu U ^\dagger \DD^\nu U \right] \cr
\lcal_3 &= - 2 i g \beta_3 \tr
\left[ \WW_{ \mu \nu } \DD^\mu U \DD^\nu U^\dagger \right] \cr
\lcal_4 &= \beta_4
\left( \tr \left[ \DD^\mu U^\dagger \DD^\nu U \right] \right)^2  \cr
\lcal_5 &= \beta_5
\left( \tr \left[ \DD^\mu U^\dagger \DD_\mu U \right] \right)^2  \cr
\lcal_6 &= - \beta_6
\tr\left[\DD_\mu U^\dagger \DD_\nu U \right]
\tr\left[\tau_3 U^\dagger \DD^\mu U\right]
\tr\left[\tau_3 U^\dagger \DD^\nu U\right]  \cr
\lcal_7 &= - \beta_7
\tr\left[\DD^\mu U^\dagger \DD_\mu U \right]
\tr\left[\tau_3 U^\dagger \DD^\nu U\right]
\tr\left[\tau_3 U^\dagger \DD_ \nu U\right]  \cr
\lcal_8 &=
\quarter g^2 \beta_8 \left( \tr \left[ U \tau^3 U^\dagger \WW_{ \mu \nu }
\right]\right)^2 \cr
\lcal_9 &= - i g \beta_9 \tr \left[ U \tau^3 U^\dagger
\WW_{ \mu \nu } \right] \tr \left[ \tau^3 \DD^\mu U^\dagger \DD^\nu U \right]
\cr
\lcal_{10} &= \half \beta_{10}
\left( \tr \left[ \tau_3 U^\dagger \DD_\mu U \right]
       \tr \left[ \tau_3 U^\dagger \DD_\nu U \right] \right)^2 \cr
\lcal_{11} &= g \beta_{11} \epsilon^{\mu\nu\rho\lambda}
 \tr \left[ \tau_3 U^\dagger \DD_\mu U \right]
 \tr \left[ \WW_{\rho\lambda} \DD_\nu U U^\dagger \right]  \cr
}
\eqn\chirall
$$
\setbox1=\vbox{
<LE>}
The numbering system is taken from [\long,\apwu].  Operators related to these
by the equations of motion have been removed.

\Appendix{C}
\titlestyle{\bf Three-field Operators}
\vskip .2truein
\message{Three-field Operators}

The effective theory is obtained from the full theory by integrating out
the heavy fields (and all heavy components of the light fields).  Before we
can do this, of course, we have to identify the heavy states.  If the
theory is written in terms of fields that are not mass eigenstates, we have
to diagonalize the quadratic part of the Lagrangian so as to write the
theory in terms of mass eigenstates; this may introduce new three-point
light vertices.  Once the Lagrangian is written in terms of mass
eigenstates, and with a suitable choice of 't~Hooft-type gauge, there are
no two-point vertices - every vertex has at least three lines.  It is,
therefore, impossible to create a tree graph with an internal heavy line
and only three external lines.  The only three-point terms in the effective
Lagrangian are those that were present explicitly in the full Lagrangian.
Any other three-point effective operators cannot get tree-level
contributions.  Of course these operators may have four, five, and
six-field parts as well, but gauge invariance guarantees that if one part
of the operator is loop level, the other parts are as well.

The situation is less straightforward once the $\su2_L \times \ui_Y$
symmetry of the effective Lagrangian is broken.  At this point $\phi
\rightarrow \phi + v/ \sqrt{2}$, and so a whole host of terms which
previously had none suddenly have three-field parts.  Are these, too,
required to come only at the loop level?  If so, that would leave the
four-fermion operators (A.7) - (A.26) as the only possible tree-level
operators.  The answer is no, not necessarily: {\it that an effective
theory and the full theory give the same Green's functions in the symmetric
phase does not imply that the two give the same Green's functions in the
broken phase.}  Nevertheless, S-matrix elements are unaffected.

To see this consider the difference between the correct and incorrect
identification of the mass eigenstates.  Let $\phi_1$ represent a light
scalar, and $\phi_2$ a heavy one.  To get the effective theory in terms of
$\phi_1$ we integrate out the heavy field $\phi_2$:
\setbox1=\vbox{ <LE>}
$$ \eqalign {
W[j_1,j_2] &= \int {\dcal\phi_1\dcal\phi_2
\exp\thinspace i \int {d^4x\left[ \lcal(\phi_1,\phi_2)
                                 + j_1 \phi_1 + j_2 \phi_2 \right]} }.  \cr
&\hbox{so} \cr
W[j_1,0] &= \int {\dcal\phi_1\dcal\phi_2 \exp\thinspace i
             \int {d^4x\left[ \lcal(\phi_1,\phi_2) + j_1 \phi_1 \right]} }  \cr
&= \int {\dcal\phi_1 \exp\thinspace i
                \int{d^4x \left[ \lcal_1 (\phi_1) + j_1 \phi_1 \right]}}
         \int {\dcal\phi_2 \exp\thinspace i
                 \int{d^4x \left[\lcal_2(\phi_1,\phi_2) \right] } }    \cr
&= \int {\dcal\phi_1 \exp i \int{ \left[ \lcal_1(\phi_1)  + j_1 \phi_1
\right]}}
                  \exp\thinspace i \int{ \lcal'(\phi_1) }    \cr
&= \int {\dcal\phi_1 \exp\thinspace i
            \int{d^4x \left[ \lcal_{eff}(\phi_1)  + j_1 \phi_1 \right]}} \cr
}
\eqn\eigen
$$
\setbox1=\vbox{
<LE>} where $\lcal(\phi_1,\phi_2) = \lcal_1(\phi_1) +
\lcal_2(\phi_1,\phi_2)$ and $\lcal_{eff} = \lcal_1 + \lcal'$.  The matching
came when we did the integral over $\dcal \phi_2$.  (We should also
integrate out the heavy components of $\phi_1$; since we're only working at
tree level here, we can neglect this.)  The light Green's functions are
just derivatives of $W[j_1,0]$ with respect to $j_1$.  Since we made no
changes in \eigen\ - we only did an integral - the light Green's functions
in the full theory (the second line) are the same as those in the effective
theory (the last line).

Now consider a theory with some mixing between $\phi_1$ and $\phi_2$.
Perhaps there is a $\phi_2 \phi_1^2$ term in the Lagrangian, so that after
spontaneous symmetry breaking ($\phi_1 \rightarrow \phi_1 + v$) a quadratic
mixing term $\phi_1 \phi_2$ appears.  The new mass eigenstates are $\phi_l
= a\phi_1 + b\phi_2$, and $\phi_h = c\phi_1 + d\phi_2$, where a, b, c, and
d are composed of numbers and coupling constants.  In this case, we should
write the theory in terms of $\phi_l$ and $\phi_h$, and integrate out the
heavy field $\phi_h$ to get the effective Lagrangian, a function of
$\phi_l$.  But let us be obstinate and instead integrate out $\phi_2$ - the
field which is not quite the heavy field.  In that case \setbox1=\vbox{
<LE>}
$$ \eqalign {
W[j_l,0] &= \int {\dcal\phi_l\dcal\phi_h
\exp\thinspace i \int {d^4x\left[ \lcal(\phi_l,\phi_h)
                                        + j_l \phi_l \right]} } \cr
W[j_l,0] &= \int {\dcal\phi_1\dcal\phi_2
\exp\thinspace i  \int {d^4x\left[ \lcal(\phi_1,\phi_2)  + j_l (a \phi_1 +
b\phi_2) \right]} }  \cr
&\hbox{now integrate out $\phi_2$ to get}  \cr
&=
\int {\dcal\phi_1\exp i \int{ \left[ \lcal_l(\phi_1) + a j_l \phi_1 \right]}}
\space \exp\thinspace i \int{ \left[ \lcal'(\phi_l) + f(b,\phi_1,j_l) \right] }
\cr
&= \int {\dcal\phi_1 \exp\thinspace i  \int{d^4x \left[ \lcal_{eff}(\phi_1)   +
a j_l \phi_1 +f(b,\phi_1,j_l) \right]}}  \cr
}
\eqn\noeigen
$$
\setbox1=\vbox{
<LE>} To get light Green's functions we differentiate with respect to
$j_l$, so $$ G^{(n)} = \bra{0} (a\phi_1 + {d f(b,\phi_1,j_l) \over dj_l})^n
\ket{0}_{j_l=0}.  $$ Again, since we've made no changes in the generating
functional, these Green's functions are identical to the Green's functions
in the full theory.  But if we don't realize that $\phi_l$ is the correct
light field, and instead couple $\phi_1$ to the source $j_l$, we don't get
the extra source coupling $f(b,\phi_1,j_1)$.  So the generating functional
we use would be incorrect - effectively we've set $a=1$ and $b=0$.  The
Green's functions we get by differentiating with respect to $j_l$ are
simpler, $$ G^{(n)} = \bra{0} (\phi_1)^n \ket{0}_{j_l=0}, $$ but they're
wrong.  The S-matrix elements, however, are identical to those produced by
the full theory, or by the correctly obtained effective Lagrangian. Setting
$a=1$ just renormalizes the Green's functions, and dropping a
source-coupling term like $f(b,\phi_1,j_1)$ changes the Green's functions,
but leaves S-matrix elements unaffected [\canlag].

In fact, we may have been making this mistake all along.  As long as we are
working in the symmetric phase, the Standard-Model fields are exactly the
light eigenstates; we diagonalize the full theory before we integrate out
the heavy fields.  But spontaneous breaking of the $\su2_L \times \ui_Y$
symmetry may change the mass eigenstates.  The fields we identified as
light and heavy in the symmetric phase may no longer be the true particle
eigenstates, and so the field coupled to the source was not what it should
have been.

In the symmetric phase, the effective Lagrangian is composed of light field
eigenstates, and we have integrated out the heavy field eigenstates.  The
full theory and the effective theory will therefore give the same Green's
functions.  Since there are no tree-level three-point Green's functions in
the full theory (other than those explicitly in the Lagrangian), there will
be no tree-level three-point Green's functions in the effective Lagrangian
(other than the Standard-Model terms already in the Lagrangian); any
operators with three-point parts must come at loop level.

In the broken phase, though, we may have integrated out the wrong fields.
It is no longer true that the full theory and the effective theory produce
the same Green's functions.  So even though there are no tree-level
three-point Green's functions in the full theory (other than those
explicitly in the Lagrangian), there may be tree-level three-point Green's
functions in the effective theory.  Effective operators which, in the
broken phase, contain three-point vertices, may still be tree-level
operators.  $\ocal_{\partial\phi}$ (A.28), for example, may have a
tree-level coefficient, even though it contains three-point terms which
(because of their dimensionality) were not in the full Lagrangian.  The
Green's functions they lead to will differ from those of the full theory,
but the S-matrix elements will be identical.

Of course if spontaneous symmetry breaking in the full theory did not mix
states, then the effective Lagrangian would give the same Green's functions
as the full theory, in both the symmetric and broken phases.  Since there
can be no tree-level diagrams in the full theory contributing to
three-point operators, none of them would have tree-level coefficients, and
so only the four-fermion operators (A.7) - (A.30) could have tree-level
couplings.  Such a case is simple to construct, but most models relevant to
the real world {\it do} have mixing, and so more than the four-fermion
operators may appear at tree level.

\vfill\eject
\refout
\vfill\eject

\bye